\documentclass[trackchanges,twocolumn]{aastex}

\let\tablenum\relax

\usepackage{natbib,epsfig,siunitx,url,graphicx,amsmath,footnote,float,xcolor,cases,lineno}

\begin{document}

\submitted{Accepted for publication in \textit{AJ}}

\title{Dynamical avenues for Mercury's origin II: in-situ formation in the inner terrestrial disk}

\author{Matthew S. Clement\altaffilmark{1}, \& John E. Chambers\altaffilmark{1}}

\altaffiltext{1}{Earth and Planets Laboratory, Carnegie Institution
for Science, 5241 Broad Branch Road, NW, Washington, DC
20015, USA}
\altaffiltext{*}{corresponding author email: mclement@carnegiescience.edu}

\begin{abstract}

Modern terrestrial planet formation models are highly successful at consistently generating planets with masses and orbits analogous to those of Earth and Venus.  In stark contrast to classic theoretical predictions and inferred demographics of multi-planet systems of rocky exoplanets, the mass ($\gtrsim$10) and orbital period ($\gtrsim$2) ratios between Venus and Earth and the neighboring Mercury and Mars are not common outcomes in numerically generated systems.  While viable solutions to the small-Mars problem are abundant in the literature, Mercury's peculiar origin remains rather mysterious.  In this paper, we investigate the possibility that Mercury formed in a mass-depleted, inner region of the terrestrial disk ($a<$ 0.5 au).  This regime is often neglected in terrestrial planet formation models because of the high computational cost of resolving hundreds of short-period objects over $\sim$100 Myr timescales.  By testing multiple disk profiles and mass distributions, we identify several promising sets of initial conditions that lead to remarkably successful analog systems.  In particular, our most successful simulations consider moderate total masses of Mercury-forming material (0.1-0.25 Earth masses).  While larger initial masses tend to yield disproportionate Mercury analogs, smaller values often inhibit the planets' formation as the entire region of material is easily accreted by Venus.  Additionally, we find that shallow surface density profiles and larger inventories of small planetesimals moderately improve the likelihood of adequately reproducing Mercury.

\end{abstract}

\section{Introduction}

\begin{figure*}
	\centering
	\includegraphics[width=.85\textwidth]{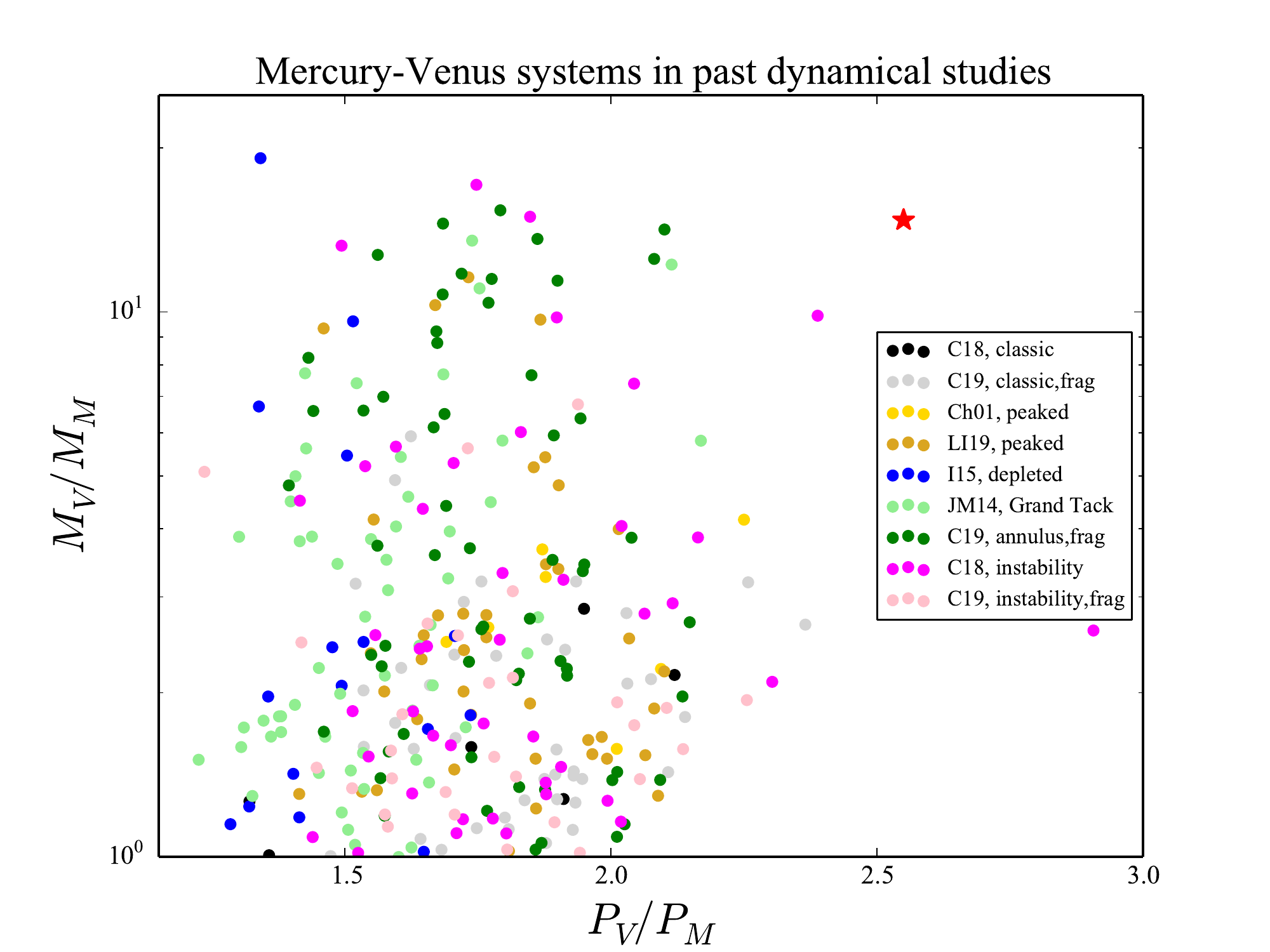}
	\caption{Summary of Mercury-Venus system mass and orbital period ratios in past dynamical studies.  For the purposes of this plot, we simply define a Mercury-Venus system as any simulation finishing with exactly two planets inside of 0.85 au.  The red star denotes the solar system value, and the different colored points correspond to simulation sets from the literature assuming different initial conditions as follows: \textbf{Black points:} A set of 100 simulations from \citet{clement18} assuming the so-called classic initial conditions \citep[$\sim$5 $M_{\oplus}$ of terrestrial planet forming material distributed uniformly between 0.5-4.0 au, e.g:][]{chambers98,ray09a}.  Here, Mercury forms directly within the massive disk in the same manner as Earth and Venus.   \textbf{Grey points:} A batch of 100 simulations from \citet{clement18_frag} assuming the identical classic initial conditions and also including a prescription \citep[section \ref{sect:methods}):][]{chambers13} accounting for the effects of collisional fragmentation.  In successful realizations, Mercury forms small as the result of a series of imperfect accretion events.  \textbf{Gold points:} A suite of 16 simulations reported in \citet{chambers01} where the Mercury forming region is modeled with a linearly increasing surface density of embryos and planetesimals between 0.3-0.7 au (hence the disk is ``peaked'').  Successful outcomes occur when Mercury forms directly from within this mass-depleted inner disk component.  \textbf{Tan points}:  40 Mercury-Venus analog systems formed in 89 simulations in \citet{lykawka19} considering similar inner disk components as \citet{chambers01}.  \textbf{Blue points}: The work of \citet{izidoro15} comprising 60 numerical simulations where the slope of the disk surface density profile was varied.  Successful Mercury analogs are produced when an embryo is scattered out of the disk.  \textbf{Green points:} A set of 125 simulations in \citet{clement18_frag} including a fragmentation prescription where the terrestrial planets form out of a narrow annulus of material ($\sim$2 $M_{\oplus}$) between 0.7-1.0 au as envisioned in \citet{hansen09}.  Successful Mercury analogs are occasionally generated when an embryo is scattered out of the annulus.  \textbf{Magenta points:} 800 simulations from \citet{clement18} where the giant planet instability \citep{Tsi05,nesvorny12} occurs within the first 10 Myr of the process terrestrial planet formation.  Appropriate Mercury analogs are produced when the giant planets' resonant perturbations liberate an embryo from the planet-forming disk and strand it on a Mercury-like orbits.  \textbf{Pink points:} A set of 600 simulations from \citet{clement18_frag} modeling the same early instability scenario and also incorporating the effects of collisional fragmentation.  The initial conditions for the terrestrial disks in both instability batches plotted here (magenta and pink points) are identical to those of the classic disk models (5 $M_{\oplus}$ of planet forming material between 0.5-4.0 au; black and grey points).}
	\label{fig:new_fig}
\end{figure*}

The latter stages of terrestrial planet formation in the solar system are thought to have unfolded as a series of giant impacts between a population of planetary-seed embryos engulfed in an ocean of smaller, $D\sim$ 1-100 km planetesimals \citep{wetherill93,weidenschilling97,koko_ida02}.  Classic studies of this scenario \citep[e.g.:][]{chambers98,chambers01,raymond04,obrien06,ray06} tend to consider a total mass of terrestrial forming material and surface density profile that is commensurate with the presumed minimum mass solar nebula \citep{mmsn,hayashi81}, and generally consistent with modern models of dust evolution in proto-planetary disks \citep[e.g.:][]{birnstiel12}.  While these types of models are successful at replicating numerous observed aspects of the modern terrestrial system, they systematically fail to generate several important features of the planet Mercury and its peculiar relationship to Venus.  We provide a summary of the types of Mercury-Venus systems formed from a variety common sets of initial conditions in the past literature in figure \ref{fig:new_fig}.

As the ultimate giant impact phase of terrestrial planet formation is highly stochastic, and unfolds over $\sim$10-100 Myr timescales \citep{earth,kleine09}, the computational cost of studying highly simplified systems of $\lesssim$1,000 objects statistically can be substantial.  Due to the small time-step required to properly resolve the innermost ($r<$ 0.5 au) region of planet-forming material, authors typically truncate the disk at $\sim$0.5 au in order to accelerate calculations \citep[e.g.:][]{jacobson14,fischer14,iz14,clement18}.  While a sharp interior edge in the distribution of solid material might seem loosely justified by pebble evaporation in the hot inner disk \citep[][we refer to the region of $a \lesssim$ 0.5 au as the inner disk throughout our manuscript]{boley14}, the preponderance of short-period planets in the exoplanet catalog suggest that planet formation in this locality is possible \citep{ray18_rev}.  However, the majority of these so-called ``hot'' planets are, perhaps, best explained by gas-driven migration of giant planet seeds \citep{izidoro17,bitsch19,lambrechts19}; a process presumably halted in the solar system by the rapid formation of Jupiter's core \citep{kruijer17}, or as the consequence of an intrinsic feature of the solar nebula such as a long-lived pressure maxima\citep{brasser20_nat}.  Nevertheless, the presumed truncation of the disk at 0.5 au is fairly arbitrary beyond the fact that it is necessary to prevent forming additional, $\sim$Earth-mass planets in the vicinity of Mercury's modern orbit \citep{chambers98}.  Indeed, authors investigating terrestrial planet formation with various approaches often explicitly indicate their intention to neglect Mercury's formation \citep[e.g.:][]{ray09a,jacobson14,levison15}.

Early attempts to directly model Mercury's accretion considered an additional region of embryos and planetesimals with a linearly decreasing surface density profile between 0.7-0.3 au \citep{chambers01,obrien06}. To first order, these models succeeded in boosting the likelihood of forming a planet near Mercury's semi-major axis interior to three larger planets.  However, the resulting Mercury analogs were too massive by around an order of magnitude (Mercury's modern mass, $M_{M}=$ 0.055 $M_{\oplus}$).  Moreover, the dynamical separation between these planets and the systems' corresponding Venus analogs (in terms of their orbital period ratio: $P_{V}/P_{M}$) were more similar to the Earth-Venus spacing ($P_{E}/P_{V,ss}=$ 1.6) than the modern Venus-Mercury period ratio ($P_{V}/P_{M,ss}=$ 2.55; see figure \ref{fig:new_fig}).  Recent work by \citet{lykawka17} and \citet{lykawka19} investigating linear profiles between 0.2-0.5 au and 0.3-0.7 au, respectively, reached the same conclusion.

It is also possible that a primordial generation of close-in embryos or even planets existed in the vicinity of Mercury's modern orbit, and were subsequently cleared out by some dynamical process.  \citet{volk15} suggested that a system of tightly-packed planets might have existed in the young solar system \citep[similar to, for example, Kepler-11 or TRAPPIST-1:][]{lissauer13,gillon17} and were eventually lost as a result of having formed in a quasi-stable configuration \citep{izidoro17}.  While connecting the solar system to the Kepler catalog in this manner is compelling, it is unclear how the modern terrestrial planets might have survived such a scenario \citep[see][for critiques of this hypothesis]{ray16,clement19_merc,lenz20,bean20}.  Additionally, \citet{ray16} proposed that the truncated initial conditions often assumed in the literature (i.e.: the absence of planetesimals and embryos interior to 0.5 au) might be generated via planetesimal shepherding if Jupiter migrated through the terrestrial-forming region during the gas disk phase \citep[as invoked in the so-called Grand Tack model:][]{walsh11,walsh16,brasser16}.  Alternatively, this classic set-up might reflect the inefficiency of planetesimal accretion in Mercury's regime of the solar system due to the presence of a primordial Si snow line \citep{morby16_ice}.  Thus, while it might be possible to roughly replicate the so-called ``classic'' initial conditions \citep[and truncate the inner disk at $\sim$0.5 au:][]{chambers98,raymond04,ray09a}, the most plausible mechanisms for accomplishing this truncation are tied to the giant planets still unconstrained migration history \citep{pierens14,ray18_rev,ribeiro20} and the unknown properties of the protoplanetary disk.  Moreover, the Mercury analogs generated from these classic initial conditions are quite scarce, and are plagued by the same issues as those formed in studies assuming an additional inner disk component (i.e.: they are systematically over-massed and too close to Venus; figure \ref{fig:new_fig}).  In \citet{clement19_merc} (henceforward, Paper I), we demonstrated this with a systematic analysis of several different proposed terrestrial evolutionary schemes \citep[][see further discussion in section \ref{sect:methods}]{walsh11,ray17sci,clement18}.  While it is possible at the $\lesssim 1 \%$ level for small ($\lesssim 0.2 M_{\oplus}$) objects with Mercury-like compositions to accrete inside of Venus' orbit, the solar system value of $P_{V}/P_{M}$ is outside of the spectrum of simulation-generated outcomes.

Of increasing interest in the recent literature, Mercury's massive iron core \citep[70-80$\%$ of its total mass:][]{hauck13,nittler17} and depleted bulk volatile content \citep[note, however that \textit{MESSENGER} inferred an unexpectedly high surface volatile inventory for the planet:][]{peplowski11,nittler11} have been interpreted to imply that its diminutive size is partially the result of an energetic, mantle-stripping collision \citep{benz88,benz07,asphaug14}.  Indeed, hydrodynamical simulations of such a scenario consistently succeed at matching Mercury's mass and core mass fraction \citep[CMF;][]{chau18}.  Moreover, recent work demonstrated that this disrupted mantle material might be easily removed via interactions with the young Sun's intense solar wind \citep{spalding20}.  However, in Paper I we found that the preferred impact geometries and velocities for such an event \citep[e.g.:][]{asphaug14,jackson18,chau18} are highly unlikely occurrences in conventional terrestrial planet formation models, though it is possible that Mercury's CMF was further altered by accreting predominantly collisionally-altered planetesimals in its late veneer \citep{hyodo20}.  While Venus might seem to be a logical target for a Mercury-forming impact, in Paper I we were unable to reproduce the solar system value of $P_{V}/P_{M}$ in simulations designed to model a proto-Mercury-Venus erosive impact.  Moreover, the fact that Venus lacks a natural satellite and internally generated magnetic dynamo has been interpreted to suggest that its growth was not interrupted by such an energetic collision \citep{jacobson17b}.  An intriguing alternative to the impact hypothesis might be tidal stripping via repeated close encounters with Venus \citep{deng20}.  While such interactions do occur in N-body simulations of terrestrial planet formation \citep{fang20}, it is not clear how Mercury's orbit would be adequately re-circularized or its semi-major axis driven away from Venus such that $P_{V}/P_{M,ss}$ is matched after such a series of encounters.  Thus, in spite of substantial effort and investigation, Mercury's enigmatic origin remains, arguably, the most significant outstanding problem in the terrestrial planet formation literature (figure \ref{fig:new_fig}).

This manuscript is part of a series of papers reexamining the Mercury problem from the ground up.  The goal of our initial few investigations is to characterize functional models for the genesis of the precise Mercury-Venus system.  In future work, we intend to more robustly develop these potentially viable scenarios within the larger context of planet formation and dynamical evolution in the solar system.  In this paper, we revisit the scenario where Mercury accretes directly from an additional, mass-depleted inner-disk component of material \citep{chambers01,obrien06,lykawka17}.  Specifically, we search for disk structures that are capable of generating the Venus-Mercury period and mass ratios with a first-order investigation of the applicable parameter space (disk surface density profile, total mass, extent and embryo-planetesimal mass distribution).  Our study is motivated by the work of \citet{iz14} and \citet{izidoro15}, where the authors varied the $a\gtrsim$ 1.0 au region of the disk's mass and structure to find the parameters that best reproduced Mars \citep[an extrapolation of the disk profile envisioned by][]{chambers02}.

\section{Methods}
\label{sect:methods}

We perform 210 N-body simulations (table \ref{table:ics}) of terrestrial planet formation with the $Mercury6$ Hybrid integrator \citep{chambers99}.  All of our simulations are integrated for 200 Myr\footnote{Note that our 200 Myr integration time is insufficient to fully characterize the long-term dynamical stability of our resultant systems.} and include algorithms designed to model the effects of collisional fragmentation \citep[the fragmentation scheme is described in greater detail in Paper I; see also:][]{leinhardt12,stewart12,chambers13,clement18_frag}.  Each computation includes Jupiter and Saturn \citep{levison03} on their presumed pre-instability orbits \citep[i.e.: in a 3:2 MMR with $a_{J}=$ 5.6 au:][]{Tsi05,nesvorny12,deienno17,clement20_instb}, use a time-step of 1.6 days, remove objects that make perihelion passages within 0.02 au of the central body \citep{chambers01}, consider objects ejected at a heliocentric distance of 15 au, and incorporate a prescription to account for the effects of relativity.  As in Paper I, the minimum fragment mass ($MFM$) in our simulations is set to 0.0055 $M_{\oplus}$ \citep[10$\%$ of Mercury's modern mass:][]{wallace17,clement18_frag}

\begin{table}
\centering
\begin{tabular}{c c c c c c}
\hline
$N_{sim}$ & $a_{in}$ (au) & $a_{out}$ (au) & $M_{tot}$ ($M_{\oplus}$) & $\alpha$ & $R$ \\
\hline
10 & 0.2 & 0.6 & 0.05 & 0.5 & 4 \\
10 & 0.2 & 0.6 & 0.05 & 1.0 & 4 \\
10 & 0.2 & 0.6 & 0.05 & 1.5 & 4 \\
10 & 0.2 & 0.6 & 0.05 & 2.0 & 4 \\
10 & 0.2 & 0.6 & 0.1 & 0.5 & 4 \\
10 & 0.2 & 0.6 & 0.1 & 1.0 & 4 \\
10 & 0.2 & 0.6 & 0.1 & 1.5 & 4 \\
10 & 0.2 & 0.6 & 0.1 & 2.0 & 4 \\
10 & 0.2 & 0.6 & 0.5 & 0.5 & 4 \\
10 & 0.2 & 0.6 & 0.5 & 1.0 & 4 \\
10 & 0.2 & 0.6 & 0.5 & 1.5 & 4 \\
10 & 0.2 & 0.6 & 0.5 & 2.0 & 4 \\
\hline
10 & 0.35 & 0.75 & 0.1 & 0.5 & 1 \\
10 & 0.35 & 0.75 & 0.25 & 0.5 & 1 \\
10 & 0.35 & 0.75 & 0.5 & 0.5 & 1 \\
10 & 0.35 & 0.75 & 0.1 & 0.5 & 4 \\
10 & 0.35 & 0.75 & 0.25 & 0.5 & 4 \\
10 & 0.35 & 0.75 & 0.5 & 0.5 & 4 \\
10 & 0.35 & 0.75 & 0.1 & 0.5 & 8 \\
10 & 0.35 & 0.75 & 0.25 & 0.5 & 8 \\
10 & 0.35 & 0.75 & 0.5 & 0.5 & 8 \\
\hline
\end{tabular}
\caption{Summary of initial conditions for our various simulation sets.  The columns are as follows: (1) the number of simulations in each set, (2) the inner disk component's inner edge, (3) outer edge, (4) total mass, (5) surface density profile power law and (6) ratio of total embryo to planetesimal mass ($R$).  In all simulations, the inner disk is comprised of 20 embryos and 200 planetesimals, while the outer disk extends from the inner disk's outer edge to 1.0 (top set of simulations) or 1.1 au (bottom set), and contains 40 embryos and 400 planetesimals with $M_{tot}=$ 2.0 $M_{\oplus}$ and $R=$ 4.}
\label{table:ics}
\end{table}

\subsection{Terrestrial disk structure}

Our work follows the example of \citet{iz14} and \citet{izidoro15} by experimenting with various total masses and surface density profiles for an additional, mass-depleted region of material.  \citet{izidoro15} studied the Mars- and asteroid belt-forming regions with a depleted outer disk.  In our case, we apply the same logic to the Mercury problem by modeling a depleted inner disk component possessing a surface density profile that increases with radial distance \citep{lykawka17}:
\begin{equation}
\Sigma(r)=
\begin{cases}
\Sigma_{in} \big(\frac{r}{1 au}\big)^{+\alpha} : \quad \textrm{inner disk} \\
\Sigma_{out} \big(\frac{r}{1 au}\big)^{-3/2} : \quad \textrm{outer disk} \\
\end{cases}
\label{eqn:alpha}
\end{equation}
where $\Sigma_{in}$ and $\Sigma_{out}$ are the disk's surface density at 1 au, and are calibrated to achieve the desired total mass for the respective disk components.  Thus, our work investigates a scenario where the terrestrial forming disk's surface density profile did not possess a sharp inner edge (discussed further in \ref{sect:motivation}, below).  In all of our simulations, the inner disk component is modeled with 20 equal-mass embryos and 200 equal-mass planetesimals.  The precise initial masses for the respective particles depend on the parameters $M_{tot}$ (the total inner disk mass) and $R$ (the ratio of total embryo to planetesimal mass: $R= M_{tot,emb}/M_{tot,pln}$); which vary in our different simulations according the values provided in table \ref{table:ics}.  In general, our Mercury-forming embryos range in mass from 0.002 to 0.02 $M_{\oplus}$, and the planetesimals in the region are assigned masses between 5.0 x $10^{-5}$ and 1.25 x $10^{-3}$ $M_{\oplus}$.  Each of the 220 inner disk particles interact gravitationally, and can experience fragmenting collisions with all other objects in the simulation (note that particles initially smaller than the $MFM$ cannot fragment until they grow larger than the $MFM$).  Thus, the main difference between embryos and planetesimals in the inner disk is their mass.  In 120 simulations, we investigate inner disks that extend from 0.2-0.6 au and possess an embryo-planetesimal ratio of $R=$ 4 \citep[based on the results of high-resolution simulations of embryo growth, e.g.:][]{walsh19,clement20_psj}, and test a range of values for $\alpha$ and $M_{tot}$ (table \ref{table:ics}).  Our selection of boundary locations for our depleted inner disk regions in these simulations are loosely based off Mercury's modern orbit, and previous works considering inner disk components spanning the region of 0.3-0.7 au that find Mercury analogs systematically form too close to Venus with such disk parameters \citep{chambers01,obrien06,lykawka17}.  Based on the most successful disk parameters from this first set of simulations (see further discussion in section \ref{sect:results}) we perform 90 additional simulations that test an inner disk component extending from 0.35 to 0.75 au.  We also vary the prescribed value of $R$ in this follow-on suite of computations.

We structure our outer disks (equation \ref{eqn:alpha}) in a manner that maximizes the probability of forming Venus, Earth and Mars analogs with the correct masses and semi-major axes.  Thus, we take the so-called ``annulus'' \citep{agnor99,morishima08,hansen09} disk conditions with a truncated outer edge (1.0 au for simulations with 0.2-0.6 au inner disks and 1.1 au for those considering 0.35-0.75 au inner disks).  In all of our simulations, the outer disk's total mass is 2.0 $M_{\oplus}$ ($R=$ 4), and is composed of 40 equal-mass embryos ($M_{emb}=$ 0.04 $M_{\oplus}$) and 400 equal-mass planetesimals ($M_{pln}=$ 0.0025 $M_{\oplus}$). Embryos in the outer disk interact gravitationally, and can experience fragmentation events with all other simulation particles.  Conversely, the 400 planetesimals cannot collide with, or feel the gravitational effects of one another.  When one of these planetesimals undergoes a fragmenting collision with an embryo or inner disk planetesimal, the resulting new fragment particles are treated as embryos.  Therefore, embryos and planetesimals in the outer disk differ in both their masses, and numerical treatment within the integration.

Our initial conditions are roughly analogous to those of the low-mass asteroid belt model \citep{iz14,izidoro15,levison15,ray17sci,ray17} or Grand Tack hypothesis \citep{walsh11,jacobson14,obrien14,walsh16,brasser16}, and do not consider the possibility that late giant planet migration sculpted the outer terrestrial disk and Mars-forming regions \citep[e.g.:][]{lykawaka13,bromley17,clement18,clement18_ab,clement18_frag}.  For recent reviews of the various proposed terrestrial evolutionary scenarios, we direct the reader to \citet{izidoro18_book_review} and \citet{ray18_rev}.

\subsection{Iron-enrichment of the Mercury-forming planetesimals}
\label{sect:motivation}

While the main motivation for our study is to reproduce the Venus-Mercury period and mass ratios in N-body terrestrial planet formation simulations, it is worthwhile to discuss our scenario in the context of the various hypotheses that aim to explain how Mercury acquired its massive iron core.  It is certainly possible that a random series of erosive impacts \citep{chambers13,clement19_merc} ensue in our mass-depleted inner disk such that Mercury finishes in a mantle-depleted state.  However, our initial conditions and proposed scenario of direct, in-situ formation are perhaps more consistent with ideas suggesting that the Mercury-forming planetesimals were already iron-enriched prior to the giant impact phase \citep[for a review of the differences between the various giant impact hypotheses and more ``orderly'' explanations for Mercury's origin in the context of \textit{MESSENGER's} findings, see:][]{ebel17}.  \textit{MESSENGER} determined that the inventories of less volatile, lithophile elements (Si, Ca, Al and Mg) in Mercury's crust are not abnormal when compared to the other terrestrial planets and chondritic compositions \citep{weider15}.  These results broadly refute ideas that silicates in the Mercury-forming region were evaporated by the intense solar activity, or that the most refractory elements preferentially condense and accrete \citep[e.g.][]{morgan80} as moderately volatile elements like Mg would be lost as well.  However, there are three promising scenarios for iron-enrichment of the material in the inner terrestrial disk that are still potentially viable:

\subsubsection{Dynamical fractionation}

\citet{wurm2013} argued that Mercury's peculiar composition might be explained by the photophoretic effect: a process through which small (e.g.: $\sim$mm-scale) particles in the gaseous disk can migrate rather substantially as the result of non-isotropic solar radiation.  The magnitude of this additional force on particles is proportional to the regions' thermal gradient, and the size of the particles themselves \citep{krauss05}.  Thus, it is possible for particles to be size-sorted through this process \citep{loesche16}.  Since the mid-plane of proto-planetary disks is thought to be highly opaque, it is unclear whether this mechanism plays a significant role in altering the chemistry of the material near Mercury's modern orbit \citep[e.g.:][]{cuzzi08}.  Nevertheless, several authors have attempted to blend dynamical and chemical models in this manner.  Notably, \citet{moritary14} used an analytical disk chemistry model that accounted for sequential elemental condensation during planetesimal formation in conjunction with N-body simulations to show that carbon-enriched planets can form at a variety of radial distances in systems with super-solar carbon abundances.  However, the authors used the same inner disk profile as \citet{chambers01}, and were thus unable to form reasonable Mercury analogs.  Additionally, \citet{pignatale16} used a two dimensional condensation model to show that two enstatite-rich regimes develop in the inner terrestrial disk, with the innermost region extending in as far as Mercury's modern semi-major axis.  While more sophisticated chemical and dynamical models are still required to fully understand whether fractionation played a significant role in altering the composition of the Mercury-forming region, such a scenario is interesting in that it does not invoke a low-probability, violent dynamical event to explain Mercury's iron content.

\subsubsection{Magnetic Aggregation}

Ferromagnetic fluids of suspended magnetized particles that behave like dipoles tend to generate chain-like structures, and the viscosity of these chains is directly related to the strength of the applied magnetic field.  As simulations \citep[e.g.:][]{dudorov14} and observations \citep[e.g.: FU Ori:][]{donati05} of proto-planetary disks indicate magnetic field strengths in the Mercury-forming regions as high as 10-100 mT, \citet{kruss18} proposed that magnetic interactions would preferentially generate large chains of iron-rich aggregates in the innermost regions of the disk \citep[see][for a similar idea]{hubbard14}.  In turn, these larger chains \citep{kruss20} would presumably be more likely to become incorporated into the planetesimal precursors to Mercury formed via gravitational collapse \citep{youdin05,johansen15,simon16}.  However, it is still unclear whether such a scenario is viable for the solar system.

\subsubsection{A carbon-rich inner disk}

\citet{ebel11} showed that mixtures of pre-solar interplanetary dust particles in  high-temperature, carbon-enriched, oxygen-depleted environments form condensates with Fe/Si ratios as high as half that of the value presumed for Mercury's bulk composition.  In particular, the effect can be quite substantial if the dust concentrated at the disk mid-plane in the Mercury-forming region and is highly enriched in anhydrous chondritic interplanetary dust particles\footnote{C-IDPs: which can possess carbon inventories an order of magnitude or so higher than that of the CI chondrites that are thought to resemble the initial composition of the solar nebula because of their similarity to the Sun's photosphere \citep[e.g.:][]{ebel02}}. If isolated from the gas, these condensates could explain the iron-enrichment of the precursors to the Mercury-forming planetesimals.  However, it is still unknown whether graphite-rich silicates were present in the innermost regions of the terrestrial disk \citep[e.g.:][]{peplowski15,vanderkaaden16}.  While this process might explain the Fe/Si content of Mercury's core and mantle (i.e.: it's bulk composition), some mechanical process would still be required to reconcile the physical size of its core (i.e.: its bulk structure and high CMF).  Thus, while we scrutinize our fully formed Mercury analogs' final CMFs in the subsequent sections assuming a chondritic distribution of planet-forming material, we note that this is not a strict constraint if the depleted, inner disk component planetesimals already possessed high Fe/Si ratios.

\subsection{Success Criteria}
\label{sect:success}

 \begin{table*}
\centering
\begin{tabular}{c c c c}
\hline
Success Criterion & Parameter & Actual Value & Accepted Value  \\
\hline
\textbf{A} & Mercury-Venus system & N/A & $N_{pln}=$2 for $a<0.85$ au; $M_{M}<M_{V}$ \\	
\textbf{B} & $M_{V}/M_{M}$ & 14.75 & $>$5.0 \\	
\textbf{C} & $P_{V}/P_{M}$ & 2.55 & $>$1.75; $q_{V}>Q_{M}$ \\	
\textbf{D} & CMF & $\sim$0.7-0.8 & $>$0.5 \\
\textbf{E} & Terrestrial system & N/A & $M_{E}$,$M_{V}>$ 0.6 $M_{\oplus}$; \\
& & & 0.05 $<M_{Ma}<$ 0.3 $M_{\oplus}$; \\
& & & $a_{E},a_{V}<$ 1.3 au; $a_{Ma}>$ 1.3 au \\
\hline	
\end{tabular}
\caption{Summary of success criteria for our simulations.  The columns are as follows: (1) the particular criterion's identifier, (2) the parameter scrutinized, (3) the solar system value and (4) the value required for a system to be successful.}
\label{table:success}
\end{table*}

In contrast to many classic studies of terrestrial planet formation, we focus our investigation almost exclusively on our simulations' ability to generate Mercury-Venus analog systems.  Thus, we do not present an analysis of our simulations' success in terms of traditional metrics related to Mars' mass, the planets' formation timescales, or the terrestrial planet systems' dynamical excitation and orbital spacing \citep[e.g.: the angular momentum deficit and radial mass concentration statistics employed throughout the literature:][]{laskar97,chambers01,ray09a} as we use initial conditions that are already well-studied and verified to be reasonably successful in this manner \citep[e.g.:][]{hansen09,walsh16,clement18_frag,lykawka20}.  Instead, we scrutinize each system against four constraints (table \ref{table:success}) that are designed to systematically compare simulated Mercury-Venus analogs with the actual pair of planets.  Additionally, we introduce a fifth constraint that evaluates the general structure of the entire terrestrial system in order to ascertain whether successfully forming Mercury-Venus pairs and Venus-Earth-Mars systems are mutually exclusive results or not.

\subsubsection{Mercury-Venus analog systems}
Criterion \textbf{A} separates systems that successfully form a Mercury-Venus duo of planets from those that do not.  Specifically, \textbf{A} requires that a simulation finish with exactly two planets (defined here as any object with $M>$ 0.01 $M_{\oplus}$) possessing semi-major axes interior to 0.85 au \citep[equidistant between Earth and Venus' semi-major axes in the modern solar system, see similar classification schemes in:][]{clement18,lykawka19} with the innermost planet less massive than the outer one.  Moreover, simulations that fail in this regard are discarded for the majority of the discussion and analysis sections of our manuscript.  Thus, only systems that form exactly two objects with $M>$ 0.01 $M_{\oplus}$, $a<$ 0.85 au, and the innermost planet less massive than the next planet are considered, regardless of the presence of additional leftover embryos and planetesimals less massive than 0.01 $M_{\oplus}$ in the region.

\subsubsection{Mass ratio}
Criterion \textbf{B} analyzes the mass ratio of the Mercury-Venus pair as defined by criterion \textbf{A}.  While many authors scrutinize terrestrial planet formation simulations by requiring that each respective planet analog finish below or above an established minimum or maximum mass tolerance \citep[typically $\lesssim$ 0.1-0.3 $M_{\oplus}$ for Mercury and Mars and $\gtrsim$ 0.5-0.8 $M_{\oplus}$ for Earth and Venus:][]{ray09a,clement18,lykawka19}, this biases ``successful'' systems towards more massive Mercury analogs and diminutive Venus's.  However, it is worth noting the obvious challenge of properly resolving Mercury's formation with initial embryo masses close to the modern mass of the planet.  For instance, the embryos in our systems testing the most massive inner disk components already possess 36$\%$ of Mercury's current mass at time zero.  To avoid over-constraining our simulations, we require that criterion \textbf{B} satisfying simulations acquire Venus-Mercury mass ratios of at least 5.0 in order to permit a small number of embryo accretion events on Mercury.  Indeed, the combination of criterion \textbf{A} and \textbf{B} adequately impose an upper mass-limit on our Mercury analogs as the most massive planet to satisfy both constraints in any of our simulations is 0.18 $M_{\oplus}$.

\subsubsection{Period ratio}
Criterion \textbf{C} considers the dynamical spacing of the Mercury-Venus system.  In the solar system, Mercury's evolution is dynamically coupled to that of Venus by strong mutual nodal forcing perturbations \citep{nobili89}.  Additionally, Mercury's proximity to the $\nu_{5}$ secular resonance with Jupiter's perihelia precession drives chaotic orbital evolution in the inner solar system \citep{laskar97,batygin15b}, and has been shown to result in possible collisional trajectories between Venus and Mercury over the expected life of the solar system \citep{laskar09}.  While, we leave the assembly of the inner solar system's secular architecture \citep[e.g.:][]{bras09} to future work, a major motivation of our study is the inability of contemporary terrestrial planet formation simulations to replicate the modern value of $P_{V}/P_{M}=$ 2.55 \citep{clement19_merc,lykawka19}.  Thus, criterion \textbf{C} requires that the final Venus-Mercury period ratio exceed 1.75.  Additionally, in Paper I we found that simulations that successfully reproduced the inner two planets' masses (criterion \textbf{B}) often possessed Mercury and Venus-analogs on crossing orbits.  Therefore, we stipulate that criterion \textbf{C} satisfying runs finish with Venus' perihelion ($q_{V}$) beyond Mercury's aphelion ($Q_{M}$).

\subsubsection{Core mass fraction}
\label{sect:cmf_calc}
As discussed in section \ref{sect:motivation}, Mercury's CMF may not represent a strict constraint for our models if the planets' high mean density is not the result of mantle removal during a giant impact \citep{ebel17}.  However, for consistency we still compute the final CMF of each Mercury analog formed in our simulations using the same procedure described in Paper I.  Criterion \textbf{D} stipulates that successful Mercury analogs attain a final CMF of at least 0.5 \citep[recall that Mercury's modern CMF is $\sim$0.7-0.8:][]{hauck13,nittler17}.  Each embryo and planetesimal in our simulations is initialized as fully differentiated with a CMF of 0.3 (30$\%$ of the total mass in an iron core and 70$\%$ represented by a silicate-rich layer of mantle material; motivated by the Earth's CMF).  Fragments are produced during the integration by dividing the mass of ejected material \citep[determined utilizing relationships from][]{leinhardt12,stewart12} into a number of equal-mass fragments masses greater than the $MFM$.  The particles are then ejected in uniformly-spaced directions within the collisional plane and velocities $\sim$5$\%$ greater than the mutual two-body escape velocity. When such an event occurs, the fragments are first generated from the mantle of the projectile, followed by its core, the target object's mantle, and finally the target's core material.  We experimented with alternative methodologies such as assigning each fragment a uniform mixture of the net ejected material.  However, we determined that the particular choice of algorithm does not qualitatively alter the final results.

\subsubsection{Score}
\label{sect:score}
In addition to evaluating each set of ten simulations based on their rates of success for each individual criterion (\textbf{A}-\textbf{D}), we assign each batch of runs a numerical score that combines the respective success rates.  To accomplish this, we linearly interpret between the solar system and simulation values of $M_{V}/M_{M}$, $P_{V}/P_{M}$, and CMF$_{M}$ to convert each system's result into a number between 0.0 and 1.0.  Period or Mass ratios less than 1.0 (i.e.: those that fail criterion \textbf{A}) and CMFs less than 0.3 receive scores of 0.0.  Similarly, results exceeding the solar system value (i.e.: $M_{V}/M_{M}>$ 14.75, $P_{V}/P_{M}>$ 2.55 and CMF $>$ 0.7) earn scores of 1.0.  As an example, consider a Mercury-Venus system that finishes with $M_{V}/M_{M}=$ 10.0, $P_{V}/P_{M}=$ 1.5, and CMF$_{M}=$ 0.3.  The system's mass score would be (10.0-1.0)/(14.75-1.0) $=$ 0.65, its orbital period score would be (1.5-1.0)/(2.55-1.0) $=$ 0.32, and the corresponding CMF score would be 0.0.  For simplicity, we then add the 30 scores for each batch of 10 simulations (table \ref{table:ics}), and multiply by 10/3 to report the final cumulative scores as a percentage.
 
 \subsubsection{Inner solar system analogs}
Finally, we investigate the correlation between the ability of our systems to form accurate Mercury analogs and their success in terms of the other three terrestrial planets.  While there are numerous ways in which we might constrain our Venus-Earth-Mars systems, we focus on the planets' respective masses and semi-major axes to avoid over-constraining our simulations.  Moreover, our selected initial disk configuration of a narrow annulus is highly successful at replicating the low orbital eccentricities and inclinations of the terrestrial planets \citep{hansen09,walsh16,clement18_frag,lykawka20}.  Therefore, we adopt a similar classification scheme to the one proposed in \citet{clement18}.  Criterion \textbf{E} requires that a system form exactly two planets with $m>$ 0.6 $M_{\oplus}$ and $a<$ 1.3 au, and exactly one Mars analog with 0.05 $<m<$ 0.3 $M_{\oplus}$ and $a>$ 1.3 au (approximately equal to Mars' modern perihelion).  This ensures that Earth and Venus are more massive than Mars by at least a factor of two.  Therefore, a system satisfying both criterion \textbf{A} and \textbf{E} necessarily contains analogs of all four terrestrial planets (albeit not necessarily an adequate Mercury analog in terms of criteria \textbf{B} and \textbf{C}).  While this scheme obviously classifies some systems that are not exact solar system analogs as successful, we find it to be an adequate prescription for analyzing how our results for Mercury depend on the properties of the other three planets.  As with our CMF calculation, we experimented with several alternative classification schemes and found that the overall trends are not particularly dependent on the the precise methodology utilized.

\section{Results}
\label{sect:results}

\begin{table*}
\centering
\begin{tabular}{c c c c c c c c c c c}
\hline
$N_{sim}$ & $a_{in}$ (au) & $a_{out}$ (au) & $M_{tot}$ ($M_{\oplus}$) & $\alpha$ & $R$ & \textbf{A} (\textbf{A+E}) & \textbf{B} (\textbf{B+E}) &\textbf{C} (\textbf{C+E}) &\textbf{D} (\textbf{D+E}) & Score\\
& & & & & & Mercury-Venus pair & $M_{V}/M_{M}$ & $P_{V}/P_{M}$ & CMF \\
\hline
10 & 0.2 & 0.6 & 0.05 & 0.5 & 4  & 3 (1) & 1 (0) & 3 (1) & 1 (1) & 17.9 \\
10 & 0.2 & 0.6 & 0.05 & 1.0 & 4  & 5 (0) & 0 (0) & 2 (0) & 2 (0) & 25.2 \\
10 & 0.2 & 0.6 & 0.05 & 1.5 & 4 & 6 (0) & 2 (0) & 4 (0) & 2 (0) & 35.8  \\
10 & 0.2 & 0.6 & 0.05 & 2.0 & 4 & 6 (1) & 1 (0) & 4 (1) & 2 (0) & 34.9 \\
10 & 0.2 & 0.6 & 0.1 & 0.5 & 4 & 5 (1) & 0 (0) & 4 (0) & 1 (0) & 27.2 \\
10 & 0.2 & 0.6 & 0.1 & 1.0 & 4 & 5 (3) & 1 (1) & 4 (2) & 0 (0) & 29.1 \\
10 & 0.2 & 0.6 & 0.1 & 1.5 & 4 & 5 (0) & 1 (0) & 3 (0) & 4 (0) & 32.5  \\
10 & 0.2 & 0.6 & 0.1 & 2.0 & 4 & 3 (1) & 0 (0) & 3 (1) & 2 (1) & 15.2  \\
10 & 0.2 & 0.6 & 0.5 & 0.5 & 4 & 0 (0) & & & & 0.0  \\
10 & 0.2 & 0.6 & 0.5 & 1.0 & 4 & 0 (0) & & & & 0.0  \\
10 & 0.2 & 0.6 & 0.5 & 1.5 & 4 & 0 (0) & & & & 0.0 \\
10 & 0.2 & 0.6 & 0.5 & 2.0 & 4 & 4 (0) & 0 (0) & 4 (0) & 2 (0) & 25.7  \\
\hline
10 & 0.35 & 0.75 & 0.1 & 0.5 & 1 & 7 (0) & 2 (0) & 5 (0) & 2 (0) & 39.8 \\\
10 & 0.35 & 0.75 & 0.25 & 0.5 & 1 & 7 (3) & 3 (1) & 4 (1) & 1 (0) & 48.0 \\
10 & 0.35 & 0.75 & 0.5 & 0.5 & 1 & 4 (1) & 1 (0) & 3 (1) & 2 (0) & 24.7 \\
10 & 0.35 & 0.75 & 0.1 & 0.5 & 4 & 6 (0) & 0 (0) & 1 (0) & 4 (0) & 39.1\\
10 & 0.35 & 0.75 & 0.25 & 0.5 & 4 & 3 (0) & 1 (0) & 3 (0) & 1 (0) & 20.0 \\
10 & 0.35 & 0.75 & 0.5 & 0.5 & 4 & 6 (0) & 0 (0) & 6 (0) & 2 (0) & 41.6 \\
10 & 0.35 & 0.75 & 0.1 & 0.5 & 8 & 4 (2) & 1 (1) & 3 (2) & 3 (1) & 23.7\\
10 & 0.35 & 0.75 & 0.25 & 0.5 & 8 & 6 (1) & 1 (1) & 3 (0) & 3 (1) & 37.5 \\
10 & 0.35 & 0.75 & 0.5 & 0.5 & 8 & 4 (1) & 0 (0) & 2 (0) & 3 (0) & 24.1 \\
\hline
\end{tabular}
\caption{Number of simulations satisfying our various success criteria (table \ref{table:success}).  The first 6 columns are the same as those of table \ref{table:ics}.  The subsequent four columns give the results for each of the success criteria: (\textbf{A}) the ability to form a Mercury-Venus pair with $a<$0.85 au and $M_{M}<M_{V}$; (\textbf{B}) a mass ratio ($M_{V}/M_{M}$) greater than 5.0 for the inner two planets; (\textbf{C}) a period ratio ($P_{V}/P_{M}$) greater than 1.75; and (\textbf{D}) a CMF for Mercury of at least 0.5.  The final column provides the simulation set's score (see \ref{sect:score} for a detailed explanation).  The values in parenthesis are the number of systems with successful Venus, Earth and Mars analogs (criterion \textbf{E}) satisfying each respective success criteria.  Note that the final four criteria can only be satisfied if criterion \textbf{A} is met.}
\label{table:results}
\end{table*}

We tabulate the number of simulations from each of our different batches (table \ref{table:ics}) that satisfy our various success criteria (table \ref{table:success}) in table \ref{table:results}.  Additionally, we denote the number of Venus-Earth-Mars analog systems (comprising a subset of 26$\%$ of all our simulations; criterion \textbf{E}) that meet each respective constraint in parentheses.  In general Mercury-Venus analog systems (criterion \textbf{A}) are fairly pervasive in the majority of our simulation batches, and several sets of initial conditions yield seven systems (of 10 total simulations) that are successful in this manner.  However, consistent with previous studies that explored narrower regions of parameter space \citep{chambers01,obrien06,lykawka17}, we find it quite difficult to reproduce the modern Venus-Mercury mass ratio (criterion \textbf{B}).  We note that the final value of $M_{V}/M_{M}$ is only moderately sensitive to the particular choice of initial inner disk mass and radial slope ($M_{tot}$ and $\alpha$; see figures \ref{fig:mtot} and \ref{fig:alpha}).  Interestingly, more massive disks tend to provide improved results when paired with a shallower inner disk slope, while the combination of a lower value of $M_{tot}$ and a larger $\alpha$ also yields reasonable success rates.  This conclusion is supported by the cumulative scores (last column of table \ref{table:results}) for the respective simulation batches.  We elaborate further on these trends in sections \ref{sect:heavy} and \ref{sect:light}.  

Our simulations indicate that criterion \textbf{C} ($P_{V}/P_{M}$) is easier to match than \textbf{B}, although the solar system value lies at the extreme of the distribution of possible outcomes for nearly all of our various disk structures.  While adequate results are produced from nearly the full range of our tested parameter space, we find that satisfactory Venus-Mercury period ratios slightly correlate with low to moderate total planetesimal masses ($R=$ 1 or 4).  We explore these trends further in section \ref{sect:r}.  

Finally, we note that our scenario provides a rather effective means of altering Mercury's final CMF; a consequence we attribute to the relatively high mutual collision velocities in the inner disk.  Indeed, all but one criterion \textbf{A} satisfying Mercury analog in our current investigation finish with CMFs that are altered from the initial value of 0.3 to some degree.  However, we find no significant trends or commonalities between our various simulations that are successful in regards to criterion \textbf{D}, and thus conclude that these systems succeed purely by happenstance \citep[i.e.: a high-CMF object coincidentally survives the simulation as the Mercury-analog, see also:][]{chambers13,clement19_merc}.  Thus, while none of our simulations simultaneously satisfy all five of our success metrics, we assess this to be the result of an over-multiplication of constraints, rather than strong mutual exclusivities between the individual criteria.  However, several simulations perform adequately when scrutinized against four of our five constraints.  We present subset of these inner solar system analogs in section \ref{sect:best}.

\subsection{Heavy vs. Light disks}
\label{sect:heavy}

\begin{figure}
	\centering
	\includegraphics[width=.49\textwidth]{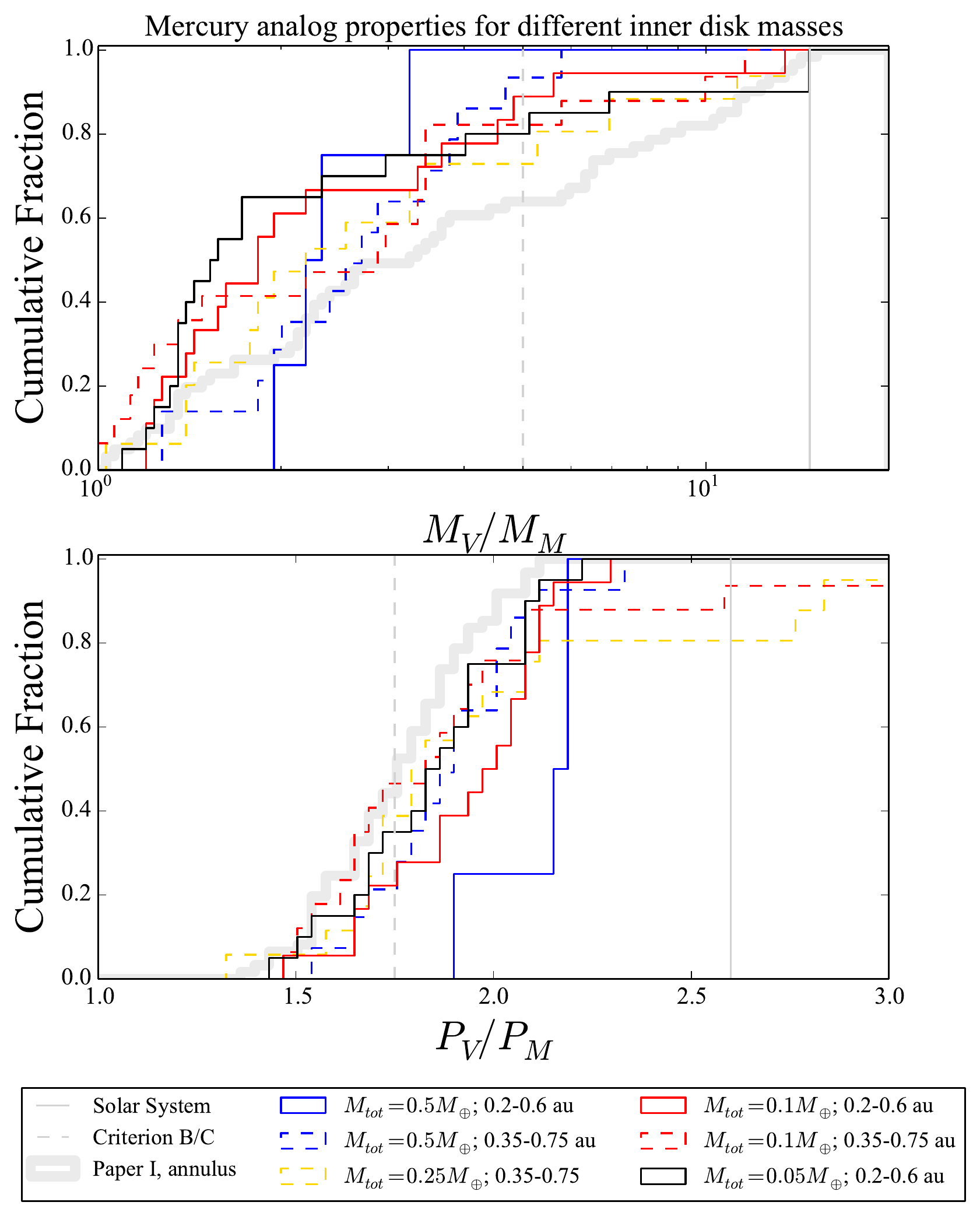}
	\caption{Cumulative distribution of Mercury-Venus system properties (top panel: mass ratio; bottom panel: period ratio) in criterion \textbf{A} satisfying simulations.  The different line colors (blue, gold, red and grey) represent the four different total inner disk masses ($M_{tot}$) employed in our simulations (table \ref{table:ics}).  Analogs originating from different inner disk radial ranges (0.2-0.6 au vs 0.35-0.75 au) are plotted with solid and dashed lines, respectively.  The vertical grey dashed lines represent the delimiting values for our success criteria \textbf{B} and \textbf{C}, while the solid grey line denotes the respective modern solar system values. The data plotted with a thick, grey, transparent line represent results from integrations presented in Paper I \citep{clement19_merc,clement18_frag} that did not include a mass-depleted, inner disk component \citep[the ``annulus'' initial conditions of][]{hansen09}.}
	\label{fig:mtot}
\end{figure}

We begin our analysis by scrutinizing the dependence of our results on the particular selection of inner disk mass ($M_{tot}$).  Figure \ref{fig:mtot} plots the cumulative fraction of final $M_{V}/M_{M}$ (top panel) and $P_{V}/P_{M}$ (bottom panel) values for the various values of $M_{tot}$ tested in our simulations.  It is important to note that we only plot and discuss systems that satisfy criterion \textbf{A} (for a measure of each system's performance that is negatively influenced by failure of \textbf{A} consult the score metric provided in table \ref{table:results}).  In this section (as well as \ref{sect:light}, \ref{sect:r} and \ref{sect:cmf}) we focus on the relative success of these Mercury-Venus pairs formed in different disk structures in terms of our other three constraints.  In sections \ref{sect:best} and \ref{sect:compare} we analyze the properties of the individual systems that are successful when measured against multiple metrics. As \textbf{A} stipulates that a system form exactly two planets with $M_{M}<M_{V}$ interior to 0.85 au, we remind the reader that a non-negligible fraction of Mercury-like planets are omitted from our analysis.  For instance, a small number of systems form more than one Mercury-analog interior to Venus.  Conversely, some simulations yield two under-massed Venus-analogs (note that neither class of simulation is considered in the majority of our analyses, specifically only figure \ref{fig:fric} compiles data from all 210 simulations).  However, the majority of the systems that fail criterion \textbf{A} are those that begin with $M_{tot}=$ 0.5, $a_{in}=$ 0.2 and $a_{out}=$ 0.6.  In these scenarios, the system begins with an excessive amount of mass concentrated interior to 0.85 au.  Typically, such simulations finish with Earth, Venus and an overly-massive Mercury analog all with $a<$ 0.85 au.  This effect is slightly lessened in our set testing the steepest inner disk mass profile ($\alpha=$ 2.0) as the more compact mass distribution tends to help Venus form closer to its modern semi-major axis.  However, the Mercury analogs in this set still tend to be over-massed.

\begin{figure}
	\centering
	\includegraphics[width=.5\textwidth]{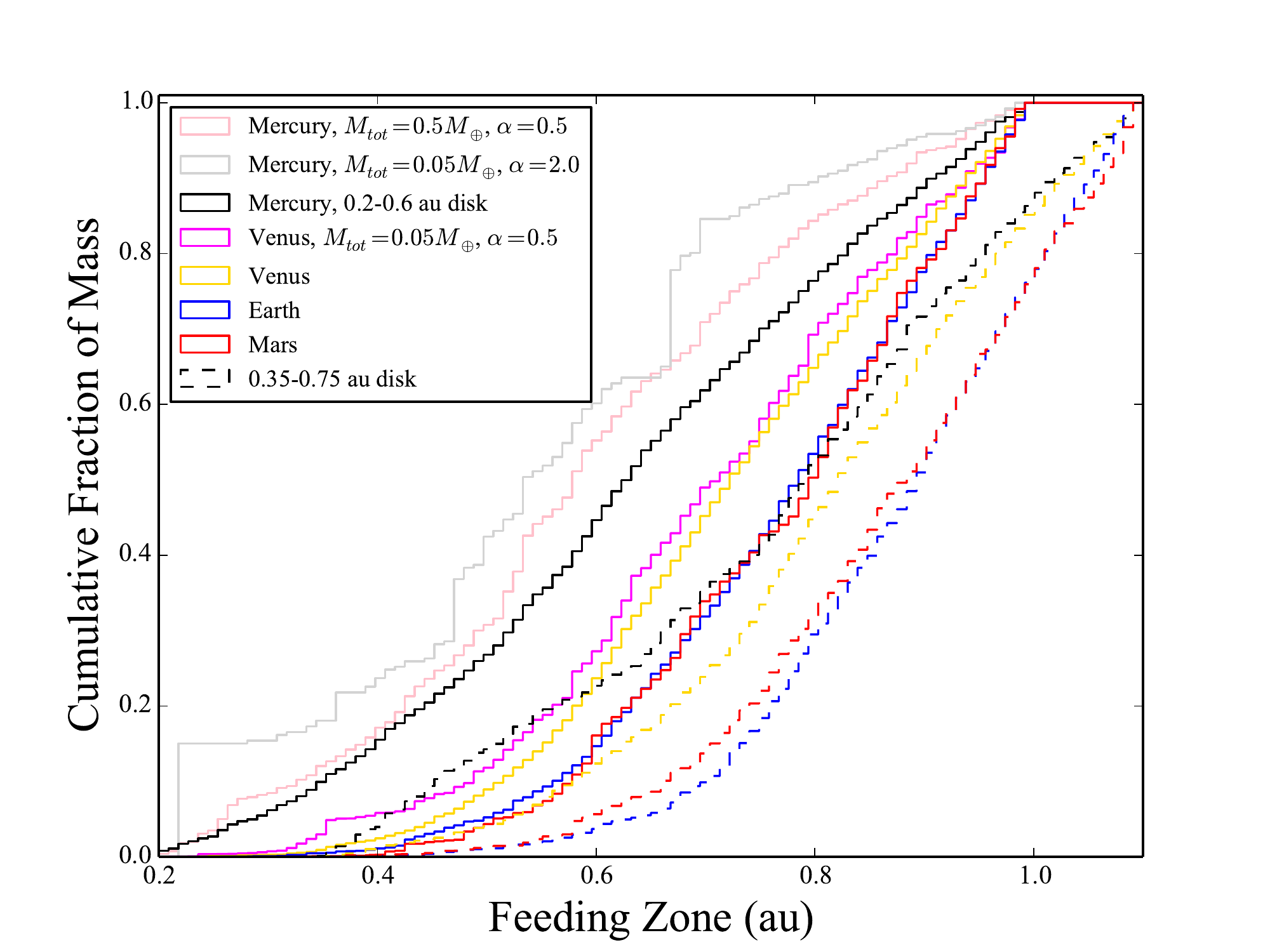}
	\caption{Feeding zones of analog planets formed in our study depicted by the mass-weighted cumulative distribution of initial semi-major axes of all chondritic (i.e. not collisional fragments) objects incorporated in to the final planets.  Analog planets are defined as follows: Mercury: 0.01 $<m<$ 0.2 $M_{\oplus}$, $a<$ 0.55 au; Venus: $m>$ 0.6 $M_{\oplus}$, $a<$ 0.85 au; Earth: $m>$ 0.6 $M_{\oplus}$, 0.85 $<a<$ 1.3 au; and Mars: 0.05 $<m<$ 0.3 $M_{\oplus}$, $a>$ 1.3 au  formed in our study.  Simulations considering inner disks extending from 0.2-0.6 au (table \ref{table:ics}) are plotted with solid lines, and runs where the inner disk was distributed between 0.35-0.75 au are plotted with dashed lines.  Mercury analogs formed in simulations testing the most extreme combinations of $M_{tot}$ and $\alpha$ are plotted separately in grey and pink for comparison.  In the same manner, Venus analogs from simulations testing the lowest value of $M_{tot}$ are plotted in magenta.}
	\label{fig:feed}
\end{figure}

To prevent Mercury, Venus and Earth from all growing in the vicinity of Venus' modern orbit, we performed an additional set of 90 simulations where the inner disk component extends from 0.35-0.75 au, and the outer section stretches between 0.75-1.1 au.  Figure \ref{fig:feed} demonstrates how this change in annulus boundaries affects the range of initial semi-major axes of embryos and planetesimals incorporated in to each final planet \citep[referred to in the subsequent text as each planet's ``feeding zone,'' e.g.:][]{kaibcowan15}.  While it is difficult to establish clear trends with only 10 simulations for each combination of varied initial conditions, it is clear from table \ref{table:results} that our most successful sets of simulations are overwhelmingly those that investigate inner disks stretching between 0.35-0.75 au.

In spite of the tendency of our more massive disks to fail criterion \textbf{A} when the inner disk boundaries are set to 0.2-0.6 au, we still assess moderate to larger values of $M_{tot}$ to be generally more successful than smaller disk masses.  While appropriate Venus-Mercury mass ratios are only produced in a small fraction of simulations, regardless of initial disk configuration (top panel of figure \ref{fig:mtot}), slightly more massive disks ($M_{tot}\simeq$ 0.1) tend to more consistently yield larger values of $P_{V}/P_{M}$.  This is a consequence of the total disk mass initially being less radially concentrated when the inner component is more massive.  In the opposite scenario, the total disk mass is more condensed towards the outer disk, thus making it easier for Venus to accrete embryos and planetesimals in the Mercury-forming region (figure \ref{fig:feed}).  When this is the case Mercury tends to remain dynamically coupled to Venus, in a sense forming as a bi-product of Venus' formation.  Contrarily, when the total inner disk mass is larger, Mercury is more likely to form independently at a smaller heliocentric distance within the inner disk.  This is also evident from the respective simulation sets' success rates for criterion \textbf{A}, \textbf{B} and \textbf{C}.  While the $M_{tot}=$ 0.05 $M_{\oplus}$ batch boasts success rates that are similar to those of the $M_{tot}=$ 0.1 $M_{\oplus}$ sets, on closer inspection we find that these ``successful'' simulations formed from less massive disks are better characterized as Venus-Earth analog systems inside of 0.85 au.  Thus, these systems satisfy criterion \textbf{A} because an Earth analog and a less massive Venus analog both finish inside of 0.85 au.  Increasing the value of $M_{tot}$ to just 0.1 $M_{\oplus}$ triples the total number of systems forming four terrestrial planets.  For these reasons, our supplementary set of 90 simulations only consider inner disk masses of 0.1, 0.25 and 0.5 $M_{\oplus}$.

Analyzing our supplementary set of simulations' (0.35-0.75 au disks) success rates in terms of criterion \textbf{B} and \textbf{C}, it is clear that low-moderate values of $M_{tot}$ tend to more consistently yield satisfactory results.  Indeed, our most successful batch of simulations in terms of our score metric consider disks with $M_{tot}=$ 0.25 $M_{\oplus}$, $\alpha=$ 0.5 and $R=$ 1.  This is not particularly surprising, given the trends discussed above.  An upper limit on the range of viable values of $M_{tot}$ is necessary to prevent Mercury from growing too large, while excessively low total masses tend to increase the chances of Venus accreting the totality of material in the inner disk.  Thus, future investigations modeling Mercury's formation from a mass-depleted interior component of the material should focus on moderate total masses (0.1-0.25 $M_{\oplus}$), and truncate the outer disks' inner edge around Venus' modern semi-major axis.

\subsection{Steep vs. Shallow disks}
\label{sect:light}

\begin{figure}
	\centering
	\includegraphics[width=.49\textwidth]{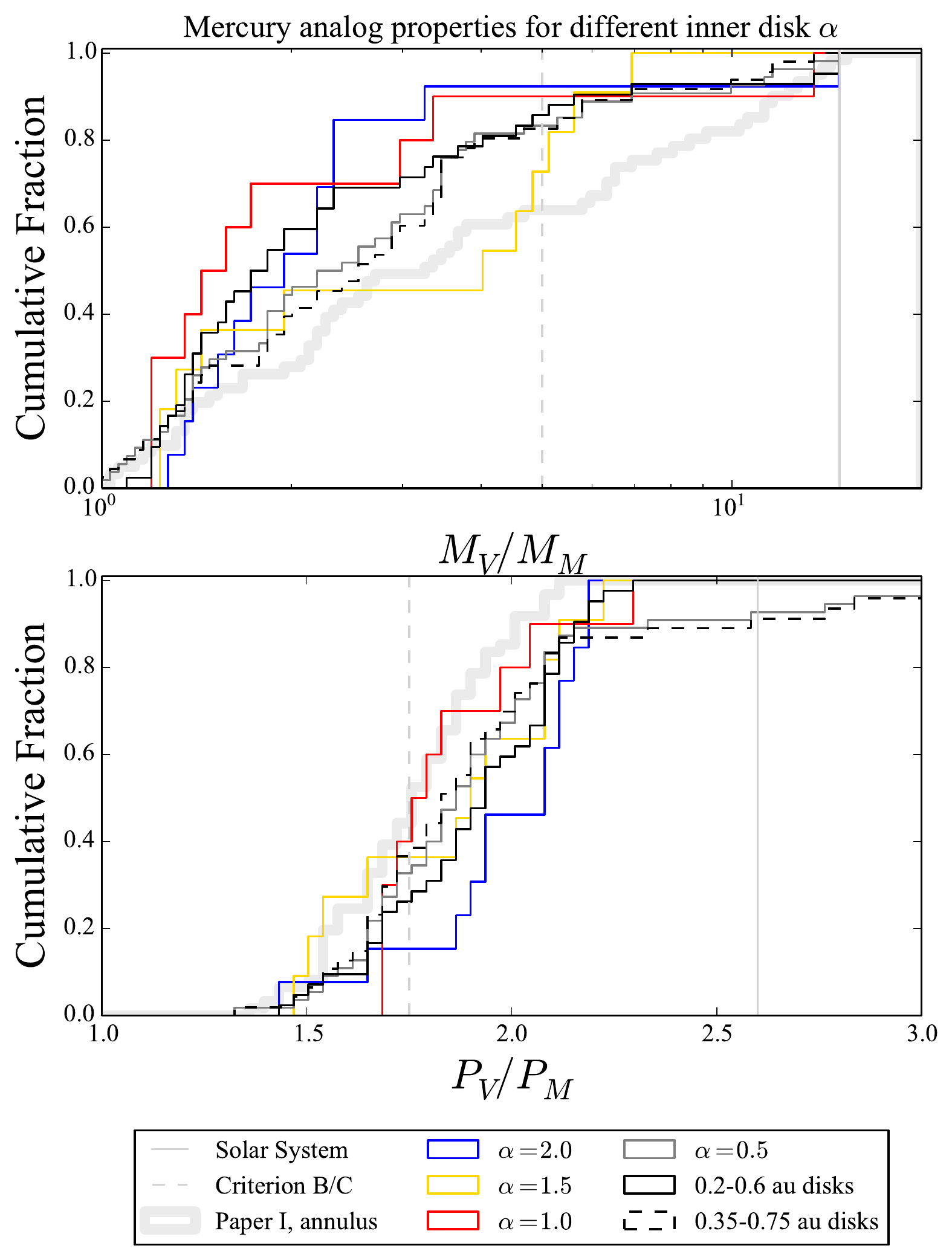}
	\caption{Same as figure \ref{fig:mtot}, except here the different line colors (blue, gold, red and grey) represent the four different inner disk slopes ($\alpha$ in equation \ref{eqn:alpha}).}
	\label{fig:alpha}
\end{figure}

Figure \ref{fig:alpha} depicts the cumulative fraction of final $M_{V}/M_{M}$ (top panel) and $P_{V}/P_{M}$ (bottom panel) ratios for the various sets of our simulations testing different inner disk surface density profiles ($\alpha$; note that only our simulations investigating 0.2-0.6 au inner disks varied this parameter).  In general, steeper slopes tend to be more successful at producing more realistic Venus-Mercury period ratios.  Conversely, we were unable to decipher any general conclusive correlations between the presumed value of $\alpha$ and the planets' final mass ratios.  Intriguingly, steeper slopes ($\alpha=$ 1.5 or 2.0) tend to be more successful for lower initial inner disk masses ($M_{tot}$), while shallower slopes are advantageous in more massive disks.  This is directly reflected in the scores for our $M_{tot}=$ 0.05 $M_{\oplus}$ (17.9, 25.2, 35.8, 34.9 in order of increasing $\alpha$) and 0.1 $M_{\oplus}$ (27.2, 29.1, 32.5, 15.2 in order of increasing $\alpha$) batches investigating $R=$ 4.  Indeed, only two simulations considering inner disk masses of 0.05 $M_{\oplus}$ simultaneously satisfy criteria \textbf{A}, \textbf{B} and \textbf{C}, and both are in the $\alpha=$ 1.5 batch.  As discussed in section \ref{sect:heavy}, our simulations investigating lower values of $M_{tot}$ systematically struggle to form Mercury analogs as Venus tends to accrete the majority of the inner disk material.  This broadening of Venus' feeding zone into the Mercury-forming region is depicted in figure \ref{fig:feed} by the difference between the solid magenta and gold lines.  However, when the inner disk surface density is sufficiently steep, the higher relative concentration of material in the Venus-forming region tends to restrict the planets' feeding zone.  This, in turn, occasionally allows Mercury to form as a stranded embryo in the inner disk component.  The clear distinction between the solid grey ($M_{tot}=$ 0.05; $\alpha=$ 2.0) and black (all simulations) lines in figure \ref{fig:feed} demonstrates Mercury's tendency to form in isolation in this manner in our integrations considering $M_{tot}=$ 0.05 $M_{\oplus}$ and $\alpha=$ 2.0.  When the initial inner disk mass is larger, Mercury forms more efficiently and in dynamical isolation from Venus with a shallower inner disk surface density profile.  This is depicted in figure \ref{fig:feed} by the difference between the solid pink ($M_{tot}=$ 0.5; $\alpha=$ 0.5) and black (all simulations) lines.  However, as these trends are rather weak, and we do not vary the initial value of $\alpha$ in our simulations investigating 0.35-0.75 au inner disk components, we conclude that the inner disk's surface density profile only mildly affects Mercury's formation.

\subsection{The effect of the total planetesimal mass ($R$)}
\label{sect:r}

\begin{figure}
	\centering
	\includegraphics[width=.5\textwidth]{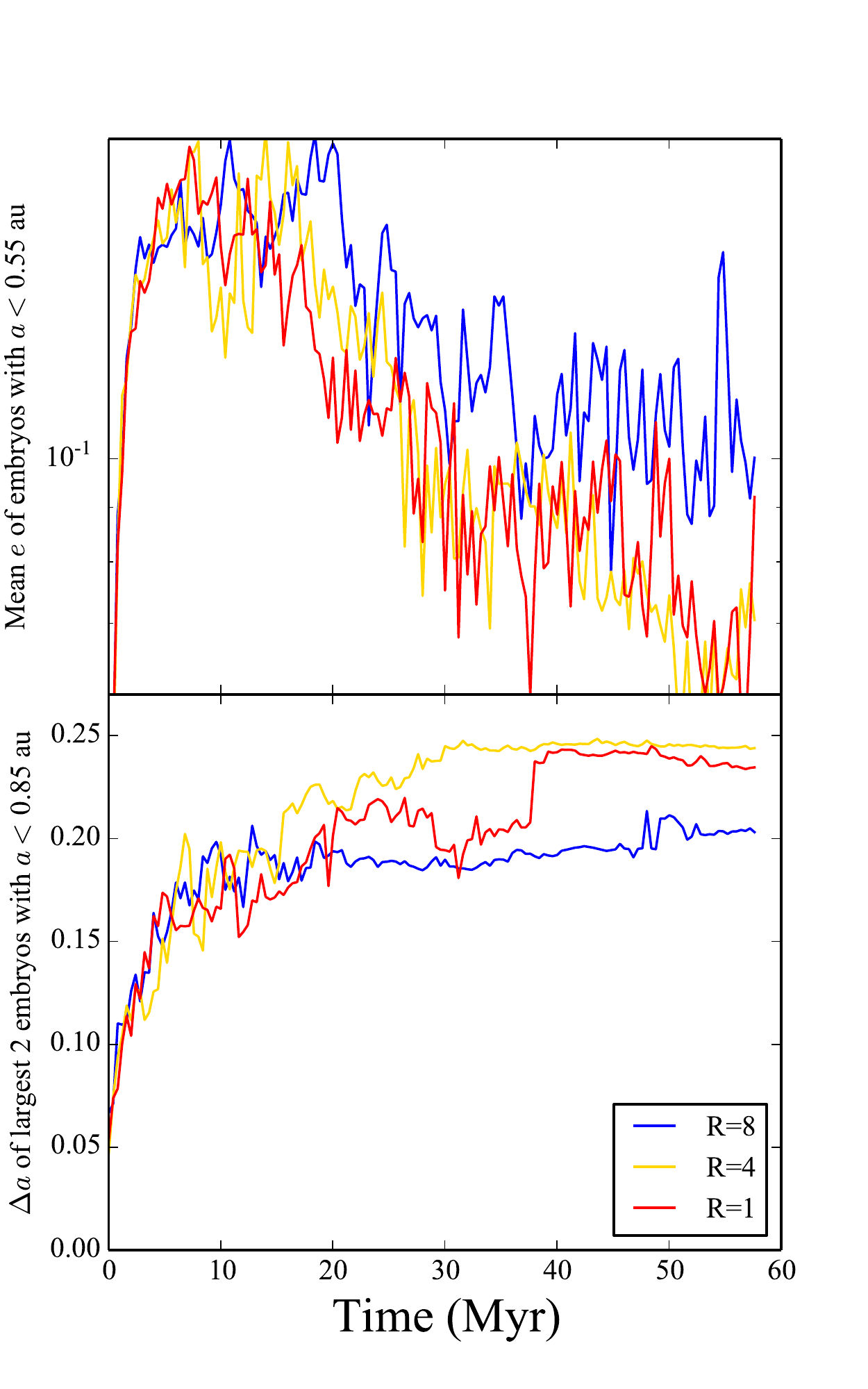}	
	\caption{Effect of the total Embryo-planetesimal mass ratio ($R= M_{tot,emb}/M_{tot,pln}$) on the evolution of Mercury and Venus in the first 60 Myr of our simulations.  The different line colors represent the 3 different values of $R$ utilized in our study (table \ref{table:ics}).  The top panel depicts the time evolution of the average eccentricity of all embryos in the Mercury-forming region ($a<$ 0.55 au).  The bottom panel shows the time evolution of the orbital spacing ($\Delta a$) of the two largest embryos with $a< 0.85$ au (ideally Mercury and Venus.  As not all of our simulations form a Mercury-Venus pair (criterion \textbf{A}), the earlier times plotted in this figure average values from all 90 of our simulations investigating 0.35-0.75 au inner disks, while the latter times only depict the 47 simulations that meet criterion \textbf{A} (table \ref{table:results}.}
	\label{fig:fric}
\end{figure}

We varied the total embryo-planetesimal mass ratio ($R= M_{tot,emb}/M_{tot,pln}$) in our additional batch of 90 simulations investigating inner disks with embryos and planetesimals distributed between 0.35-0.75 au.  Naively, one might expect the additional dynamical friction \citep[e.g.:][]{obrien06,ray06,jacobson14,lykawka19} from a more massive swarm of planetesimals to restrict the feeding zones of each planet, but we find this effect to be almost negligible.  However, we note that our runs investigating the most extreme bimodal mass distribution ($R=$ 8) tend to struggle to satisfy criterion \textbf{C} (table \ref{table:results}) and adequately reproduce the Venus-Mercury period ratio.  We assess this to be a direct consequence of the systematically weaker dynamical interactions between the growing embryos and smaller planetesimals (though it is difficult to decouple this trend from the effects of varying $M_{tot}$ in the final batch scores).  In the top panel of figure \ref{fig:fric} we show the mean eccentricity of all embryos in the Mercury-forming region ($a<$ 0.55 au) in our three batches of simulations investigating different values of $R$.  While the difference between the $R=$1 and 4 sets is minor, the embryos in the $R=$ 8 runs possess substantially hotter eccentricity distributions for the majority of Mercury's growth timescale (the average time to reach 90$\%$ of its final mass: $\lesssim$ 50 Myr for all of our different simulation sets).  As the embryos attain higher eccentricities in this manner, they interact more strongly with the growing Venus, and are thus more likely to eventually be incorporated into Venus.  Contrarily, when the embryo's eccentricity distribution is colder, the innermost embryos are dynamically sequestered from Venus in a manner that allows them to combine and accrete to form a Mercury analog in isolation.  Through this process, the final planets tend to possess larger radial offsets from Venus, as demonstrated in the bottom panel of figure \ref{fig:fric}.

It is worth pointing out that the more extreme bimodal distributions of material we find to be less successful are, perhaps, more consistent with studies of runaway growth \citep{koko_ida_96} during the gas disk phase.  Indeed, recent high-resolution investigations of planetesimal collisional evolution increasingly find a strong radial dependence to the efficiency of runaway growth \citep{carter15,walsh19,clement20_psj,woo21}.  Thus, we acknowledge that our $R=$ 1 disk might not be an accurate representation of the potential disk conditions in the Mercury forming region.  However, as our $R=$ 4 disks still yield reasonable results (figure \ref{fig:fric} and table \ref{table:results}) we do not assess this to be a serious shortcoming of our model.  Moreover, as the precise orbital and size distributions of the $r<$ 0.5 au section of the solar system's terrestrial-forming disk remain unconstrained, we plan to further investigate the feasibility of our proposed initial conditions with high-resolution disk models in future work.  It seems reasonable to speculate that a population of planetesimals and embryos drifting inward via Type I migration might be stranded in the Mercury-forming region during the disk's photo-evaporation phase.  Through this process, it might be possible for a low-mass inner component of the terrestrial disk to develop an over-abundance of planetesimals relative to larger embryos.

\subsection{Eroding Mercury's mantle}
\label{sect:cmf}

\begin{figure}
	\centering
	\includegraphics[width=.5\textwidth]{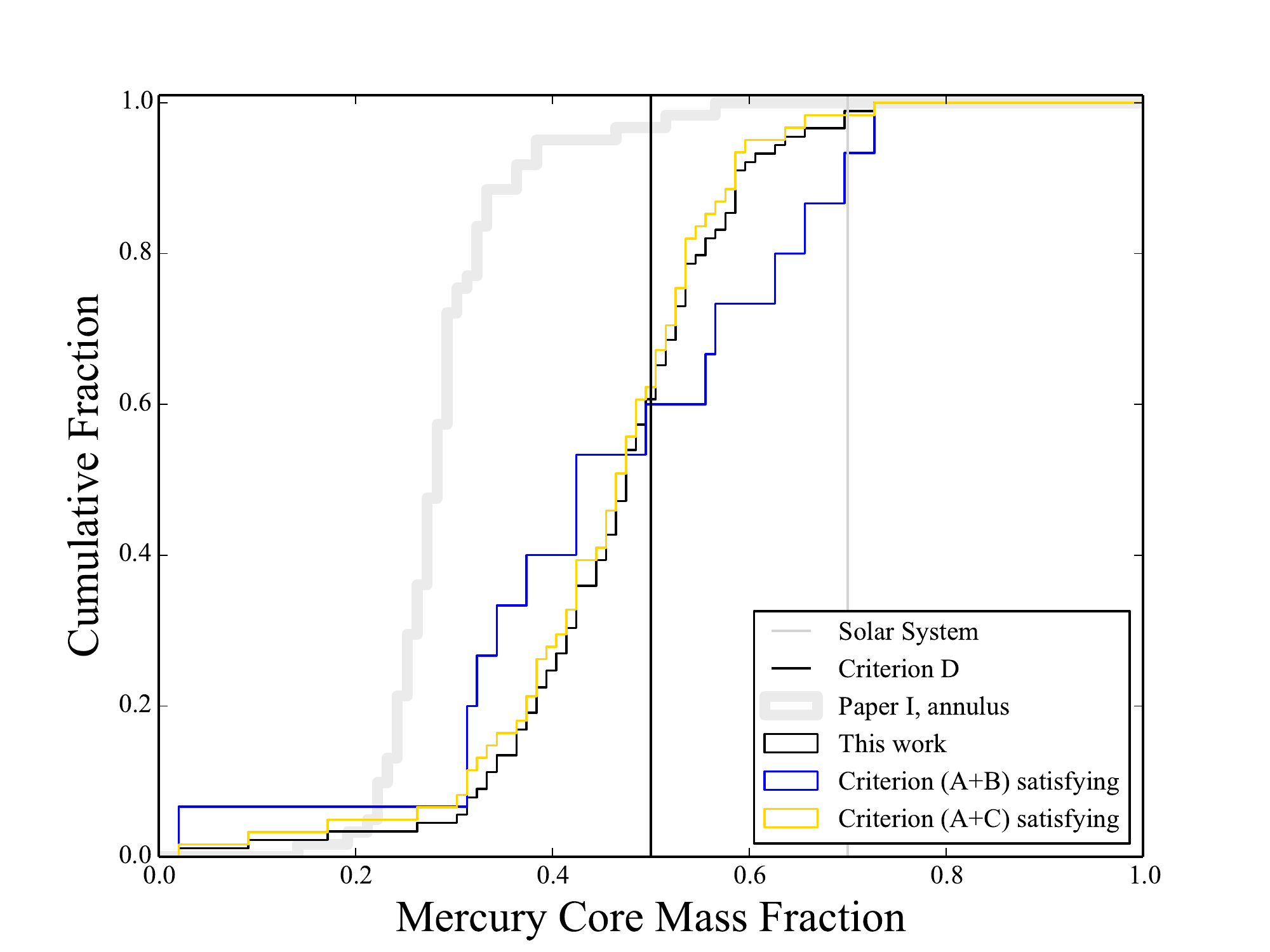}
	\caption{Cumulative distribution of Mercury analog CMFs in systems satisfying criterion \textbf{A} for all simulations presented in this work (black line), those satisfying both \textbf{A} and \textbf{B} (blue line), analogs successful in terms of both \textbf{A} and \textbf{C} (gold line), and comparison integrations from Paper I \citep{clement19_merc,clement18_frag} that did not include a mass-depleted, inner disk component \citep[the ``annulus'' initial conditions of][]{hansen09}.  Recall that the fragmentation algorithm, $MFM$ setting, and CMF calculation utilized here are identical to the methodology employed in Paper I.}
	\label{fig:cmf}
\end{figure}

While we argue that Mercury's final CMF might not represent a strict constrain for our models (section \ref{sect:motivation}), the collisional environment in our systems' inner disks provides an efficient mechanism for altering the planets' bulk composition.  Indeed, $\sim$ 40$\%$ of the Mercury analogs in our criterion \textbf{A} satisfying systems possess final CMFs in excess of 0.5 (criterion \textbf{D}).  Figure \ref{fig:cmf} plots the cumulative distribution of Mercury analog CMFs, compared to a similar set of simulations taking the same annulus initial conditions \citep[without an inner disk component:][]{hansen09} from Paper I.  It is not surprising that the Mercury-like planets generated in these simulations tend to have CMFs that concentrate around 0.3, as the majority of these planets form when an embryo is scattered from the annulus onto a relatively isolated orbit.  Thus, the final analogs are far less compositionally processed than those produced in this manuscript from an inner disk of embryos and planetesimals.  It is clear, however, that the iron contents of our contemporary simulations' Mercury analogs (independent of their success in terms of criteria \textbf{B} or \textbf{C}) are systematically enriched compared to that of their initial embryo precursors.  Indeed, 100$\%$ of our final Mercury analogs experience at least one imperfect collision over the duration of our simulations.  Curiously, while the ejected mantle material is predominantly removed from our systems via mergers with the Sun, it does so rather indirectly.  As our terrestrial-forming disks extend over a more expansive radial range than those in Paper I, collisional fragments are far more likely to be transferred to different embryos rather than be absorbed by the target embryo or central body.  Indeed, fragments produced in Paper I are $\sim$320$\%$ more likely to be removed from the simulation via collision with the Sun, and $\sim$14$\%$ less likely to be accreted by an embryo other than original target particle.  However, \textit{embryos} in our simulations considering an inner disk are nearly \textit{twice} as likely to be removed via merger with the central star.  Thus, while the ultimate fate of the mantle material removed from our Mercury analogs is the Sun, the material tends to be transferred to different embryos within the inner disk before eventually being expelled from the system.

\subsection{Preferred disk structure}
\label{sect:best}

\begin{figure*}
	\centering
	\includegraphics[width=.8\textwidth]{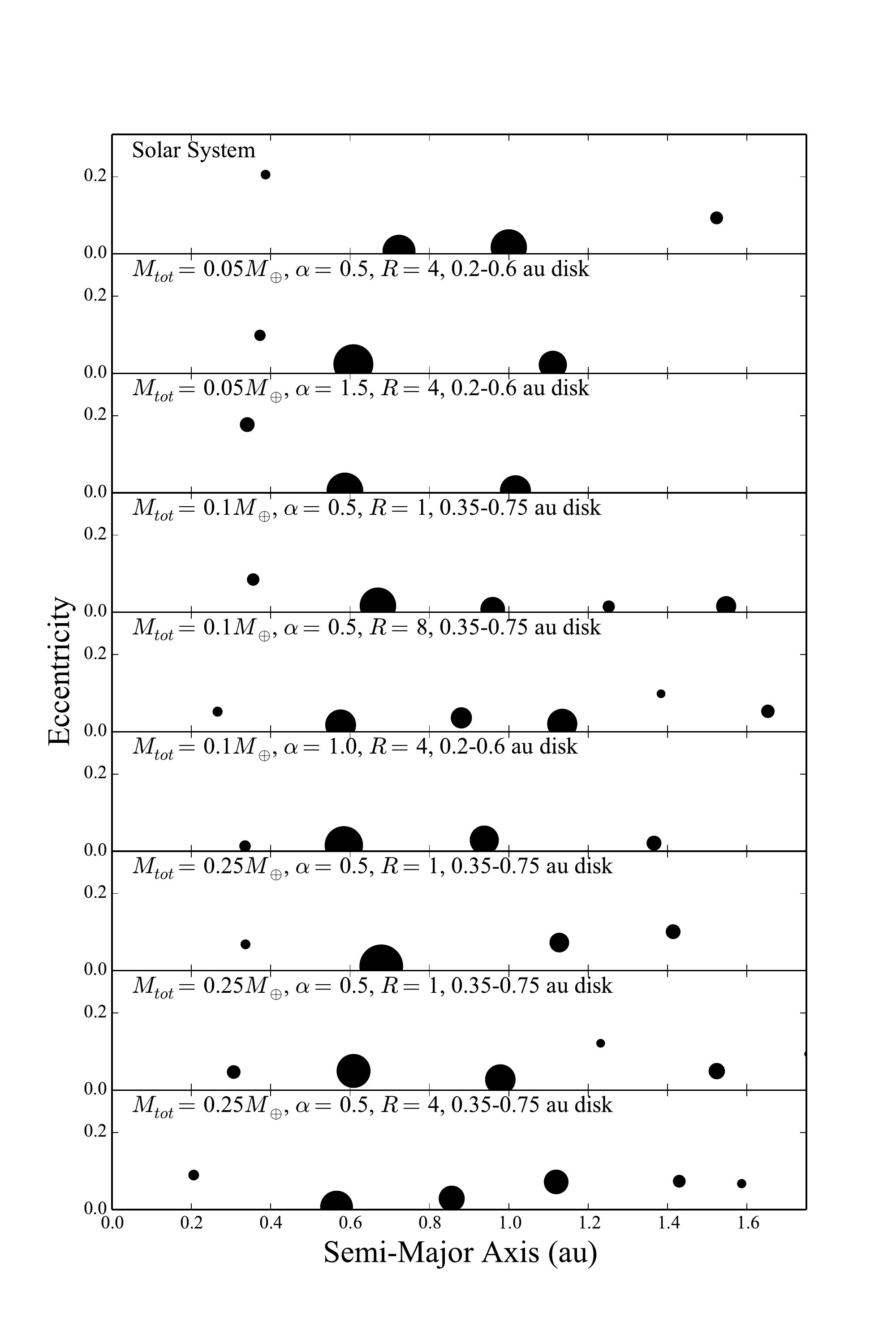}
	\caption{Semi-major Axis/Eccentricity plots of eight systems forming successful Mercury-Venus pairs (in terms of criteria \textbf{A}, \textbf{B} and \textbf{C}) from our investigation.  The size of each point corresponds to the mass of the particle.  The solar system's terrestrial planets are plotted in the top panel for comparison.}
	\label{fig:combined}
\end{figure*}

\begin{figure}
	\centering
	\includegraphics[width=.5\textwidth]{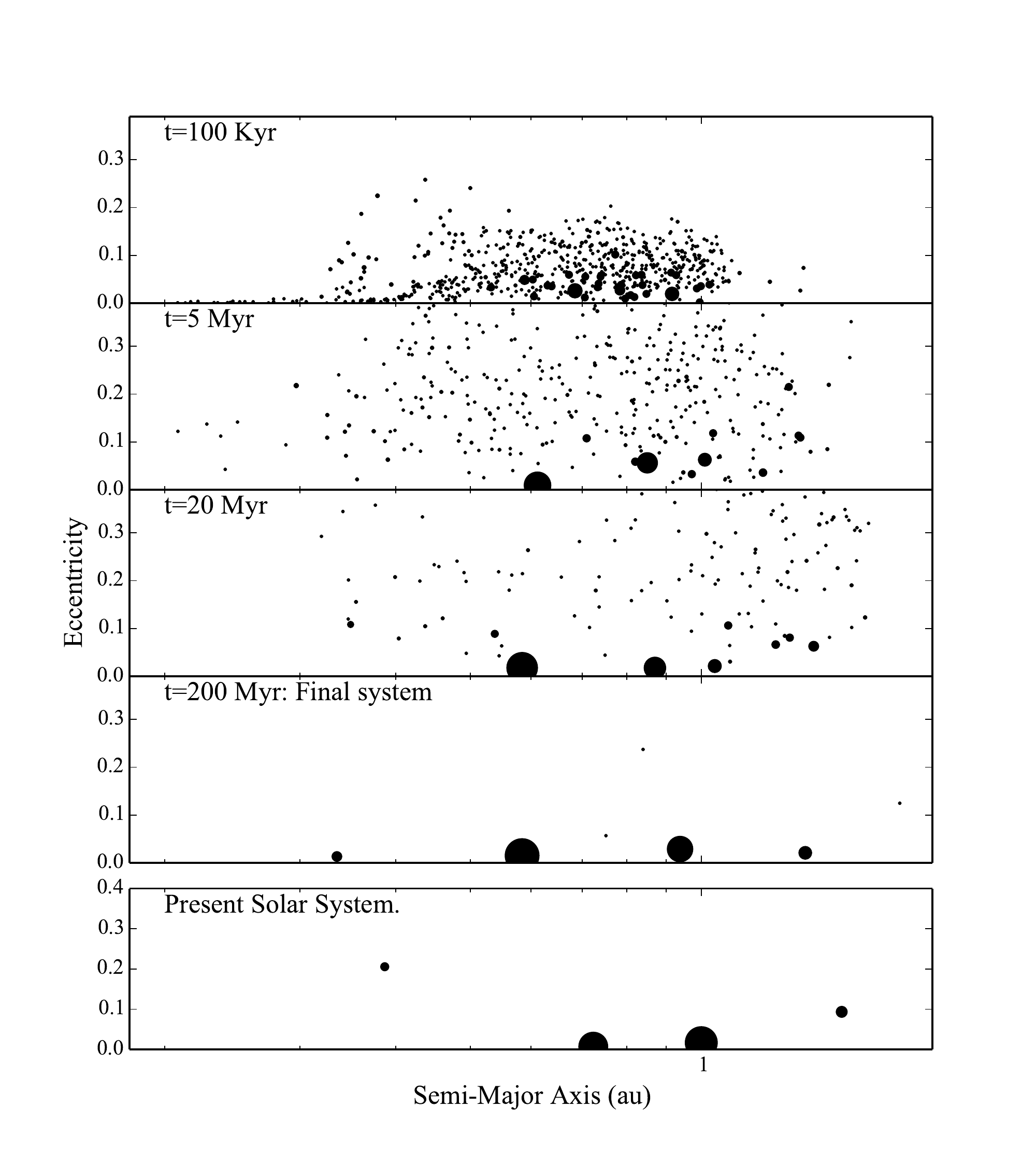}
	\caption{Semi-Major Axis/Eccentricity plot depicting the evolution of a successful system with $a_{in}=$ 0.2 au, $a_{out}=$ 0.6 au, $M_{tot}=$ 0.1 $M_{\oplus}$, $\alpha=$ 1.0 and $R=$ 4. The size of each point corresponds to the mass of the particle. The final planet masses are 0.083, 1.12, 0.63 and 0.15 $M_{\oplus}$, respectively.  Note that two small, additional Mars analogs ($M=$ 0.04 and 0.02 $M_{\oplus}$) surviving exterior to Mars ($a>$ 1.75 au) are not depicted in panel 4.}
	\label{fig:fig2}	
\end{figure}

Our simulations lead us to favor an inner disk component of moderate total mass, shallow slope, and low to moderate $R$.  Additionally, we find that mass depletion interior to 0.75 au (rather than 0.6 au) tends to boost the probability of forming a successful Mercury-Venus system.  A sample of eight systems that simultaneously satisfy criteria \textbf{A}, \textbf{B} and \textbf{C} is plotted in figure \ref{fig:combined}.  An example of a successful evolution of one of these systems (the one depicted in panel six of figure \ref{fig:combined}) is plotted in figure \ref{fig:fig2}.  The system begins with $M_{tot}=$ 0.1 $M_{\oplus}$, $\alpha=$ 1.0 and $R=$ 4.  Venus and Earth grow rapidly from seed embryos near the center of the disk at $a_{V,o}=$ 0.69 and $a_{E,o}=$ 0.87 au.  By $t=$ 2.8 Myr Venus attains half its ultimate mass.  Similarly, Earth grows to 50$\%$ of its final size in 4.8 Myr.  However, the evolution of the two larger planets subsequently bifurcate as Venus continues to rapidly accrete embryos and planetesimals from both the outer and inner disk in a manner such that it attains 80$\%$ of its eventual mass at $t=$ 12 Myr.  Conversely, Earth slowly grows to $\sim$65$\%$ of its ultimate size before experiencing a final giant impact with a 0.19 $M_{\oplus}$ proto-planet originally seeded at 0.97 au at $t=$ 25.4 Myr \citep[we note that these divergent accretion histories are a reasonable example of the scenario proposed by][that aims to explain Venus' lack of an internally generated magnetic field and natural satellite]{jacobson17b}.

Curiously, the Mercury analog in figure \ref{fig:fig2} originates when a mantle-only fragment at $r=$ 0.45 au produced in a previous event collides with an embryo initially positioned at $a=$ 0.56 au at $t=$ 10.0 Myr.  The collision yields three remnant particles with roughly equal masses of $\sim$0.005 $M_{\oplus}$.  Interestingly, one of these remnants is composed entirely of mantle material, another is totally derived from the core of the target embryo, and the final fragment (that goes on to become the Mercury analog) is constructed from a combination of the mantle-only projectile, the embryo's mantle, and part of the embryo's core.  However, as the Mercury analog only possesses $\sim$ 6$\%$ of its ultimate mass at this point, the random assignment of material by our algorithm (section \ref{sect:cmf_calc}) has little effect on the planets' ultimate composition.  The fragment continues to slowly accrete planetesimals and embryos, thereby increasing its total mass to 0.042 $M_{\oplus}$ (half its final value) by $t=$ 20.0 Myr.  Over the subsequent 7 Myr the eventual Mercury analog experiences a series of 7 giant impacts with embryos in the inner disk which cumulatively boost its mass to 0.067 $M_{\oplus}$.  The planets' mass remains largely unchanged throughout the next $\sim$30 Myr aside from a few impacts with diminutive planetesimals, and three minor fragmenting collisions which erode a small amount of mantle material.  At $t=$ 86.6 Myr Mercury experiences it's final giant impact with a 0.009 $M_{\oplus}$ embryo.  Thus, the Mercury analog in our example simulation has the longest accretion timescale of the four final terrestrial planets.  Through this complex formative epoch, the planet attains a final mass of 0.083 $M_{\oplus}$ ($M_{V}/M_{M}=$ 13.5), semi-major axis of 0.33 au ($P_{V}/P_{M}=$ 2.33)  and a CMF of 0.32.

It is worth mentioning that this simulation's Mercury analog's orbit is not nearly as dynamically excited ($e=$ 0.013; $i=$ 2.0$\degr$) as in the actual solar system ($e_{M}=$ 0.21; $i_{M}=$ 7.0$\degr$).  This is also the case for the vast majority of the other Mercury-like planets generated in our study.  Figure \ref{fig:mercs} plots the orbits of our various criterion \textbf{A} satisfying Mercury analogs in $a/e$, $a/i$ and $a/m$ space; with the systems that are successful in terms of both \textbf{A} and \textbf{B} isolated in the right panel of the plot.  It is clear from this figure and figure \ref{fig:combined} that even the Mercury analogs with masses most akin to that of the real planet tend to possess dynamically cold orbits.  However, a lower primordial eccentricity and inclination is likely advantageous as Mercury's orbit is easily excited during the giant planet instability \citep{Tsi05,nesvorny12}.  Moreover, \citet{roig16} found that Mercury's precise orbit is reasonably explained by dynamical perturbations from Jupiter's step-wise semi-major axis evolution on an initially circular, co-planar Mercury \citep[the so-called ``jump:''][]{bras09,nesvorny13}.  Thus, we conclude that the tendency of our scenario to yield dynamically cold Mercury analogs is not particularly concerning as our simulations do not consider giant planet migration.  However, it is unclear whether an early dynamical instability \citep[transpiring in conjunction with the giant impact phase of terrestrial planet formation as proposed in][]{clement18} would be compatible with our scenario and adequately excite Mercury's eccentricity and inclination.

\subsection{Comparison with Paper I}
\label{sect:compare}

\begin{figure*}
	\centering
	\includegraphics[width=.49\textwidth]{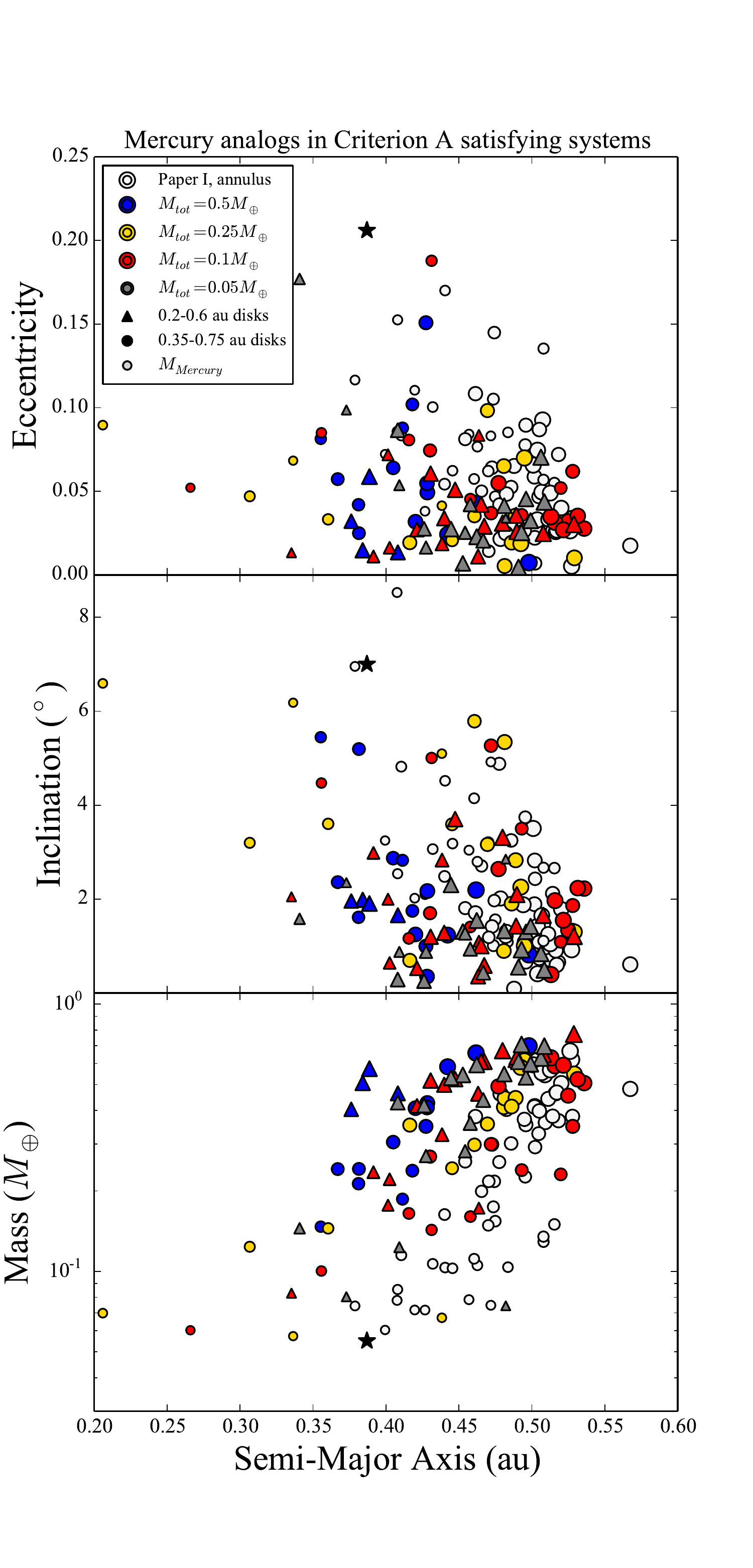}
	\includegraphics[width=.49\textwidth]{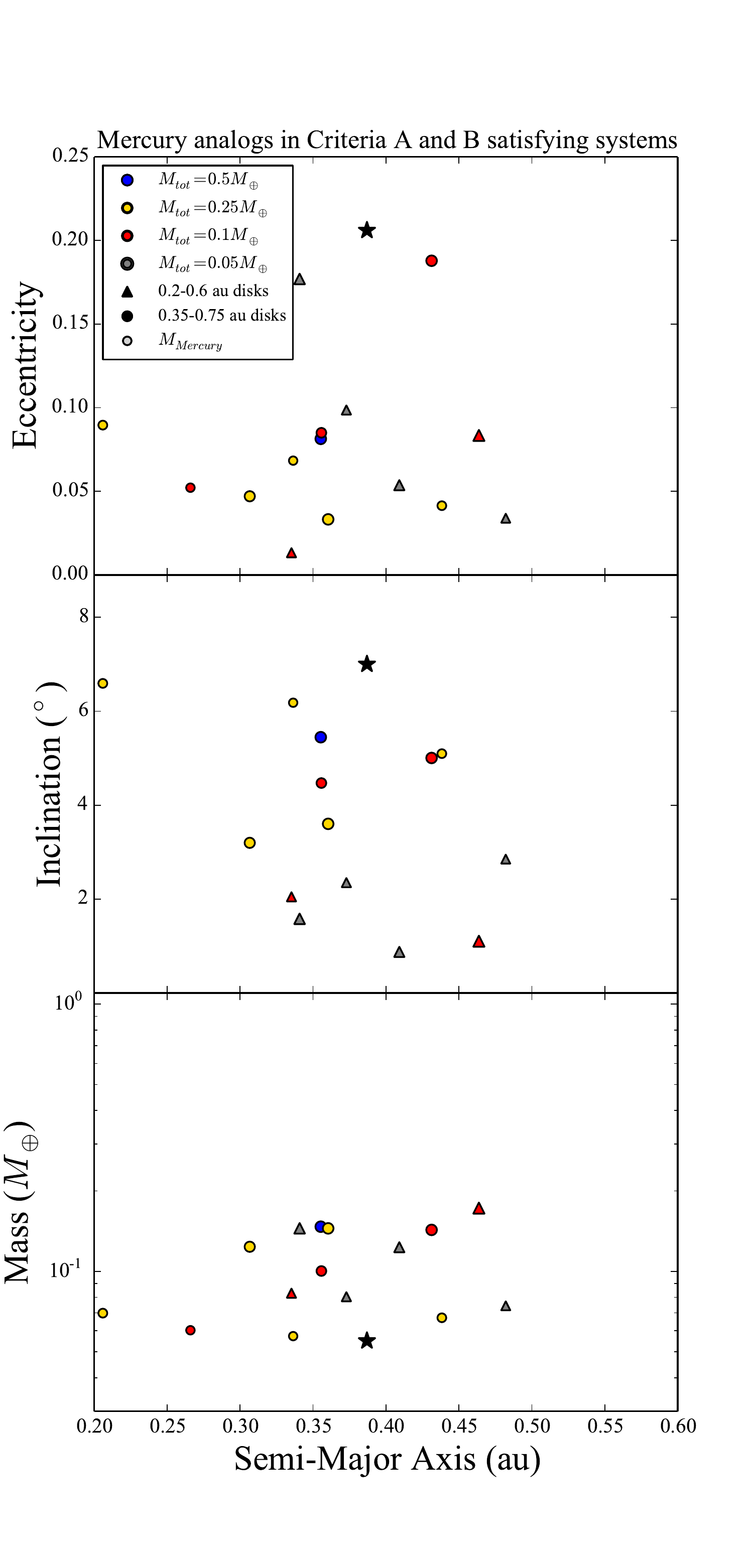}
	\caption{Final orbits in our various batches of simulations (table \ref{table:ics}) that varied the total inner disk mass ($M_{tot}$: blue, gold, red and green points) for systems satisfying criterion \textbf{A} (left panel), and both \textbf{A} and \textbf{B} (right panel).  For reference, Mercury analogs originating from different inner disk radial ranges (0.2-0.6 au vs 0.35-0.75 au) are plotted with different symbols.  The top panels plot eccentricity vs. semi-major axis for each Mercury analog, the middle sub-panels display the same systems in $a/i$ space, and the bottom panels plot the data in $a/m$ space.  The size of each point is proportional to the analog's mass (the mass of Mercury is plotted in grey in the upper left corner of each panel for reference).  The data plotted in light grey represent results from integrations presented in Paper I \citep{clement19_merc,clement18_frag} that did not include a mass-depleted, inner disk component \citep[the ``annulus'' initial conditions of][]{hansen09}.}
	\label{fig:mercs}
\end{figure*}

The left panel of figure \ref{fig:mercs} compares the orbits of the criterion \textbf{A} satisfying Mercury analogs generated in this manuscript to a set presented in Paper I (note the fragmentation algorithm and $MFM$ setting used in this study are the same as in Paper I).  As this panel compiles many systems that are unsuccessful in terms of matching Mercury's mass, we isolate the 15 simulations (table \ref{table:results}) that satisfy both criteria \textbf{A} and \textbf{B} in the right panel of figure \ref{fig:mercs}.   In Paper I we noted that a narrow annulus \citep{hansen09} yielded larger values of $P_{V}/P_{M}$ and $M_{V}/M_{M}$ (though not as large as those of the real system) than the classic, extended disk model of \citet{chambers98} perturbed by the giant planet instability as envisioned in \citet{clement18}.  As we have reiterated throughout our manuscript, our new simulations' ability to form adequate Mercury-Venus analogs represents a marked improvement from the annulus models of Paper I.  While the properties of the planets themselves are not particularly different (figures \ref{fig:mtot}, \ref{fig:alpha} and \ref{fig:mercs}), our new models substantially improve the \textit{rate} of generating Mercury-Venus systems of appropriate mass and orbital spacing.  Specifically, 20$\%$ of our most successful set of initial conditions (0.35-0.75 au disk, $M_{tot}=$ 0.1 $M_{\oplus}$, $\alpha=$ 0.5, $R=$ 1) and 10$\%$ of all our 0.35-0.75 au disk simulations simultaneously satisfy criteria \textbf{A}, \textbf{B} and \textbf{C}.  Conversely, only 5.6$\%$ of the annulus models from Paper I were successful in this regard.

\begin{figure}
	\centering
	\includegraphics[width=.5\textwidth]{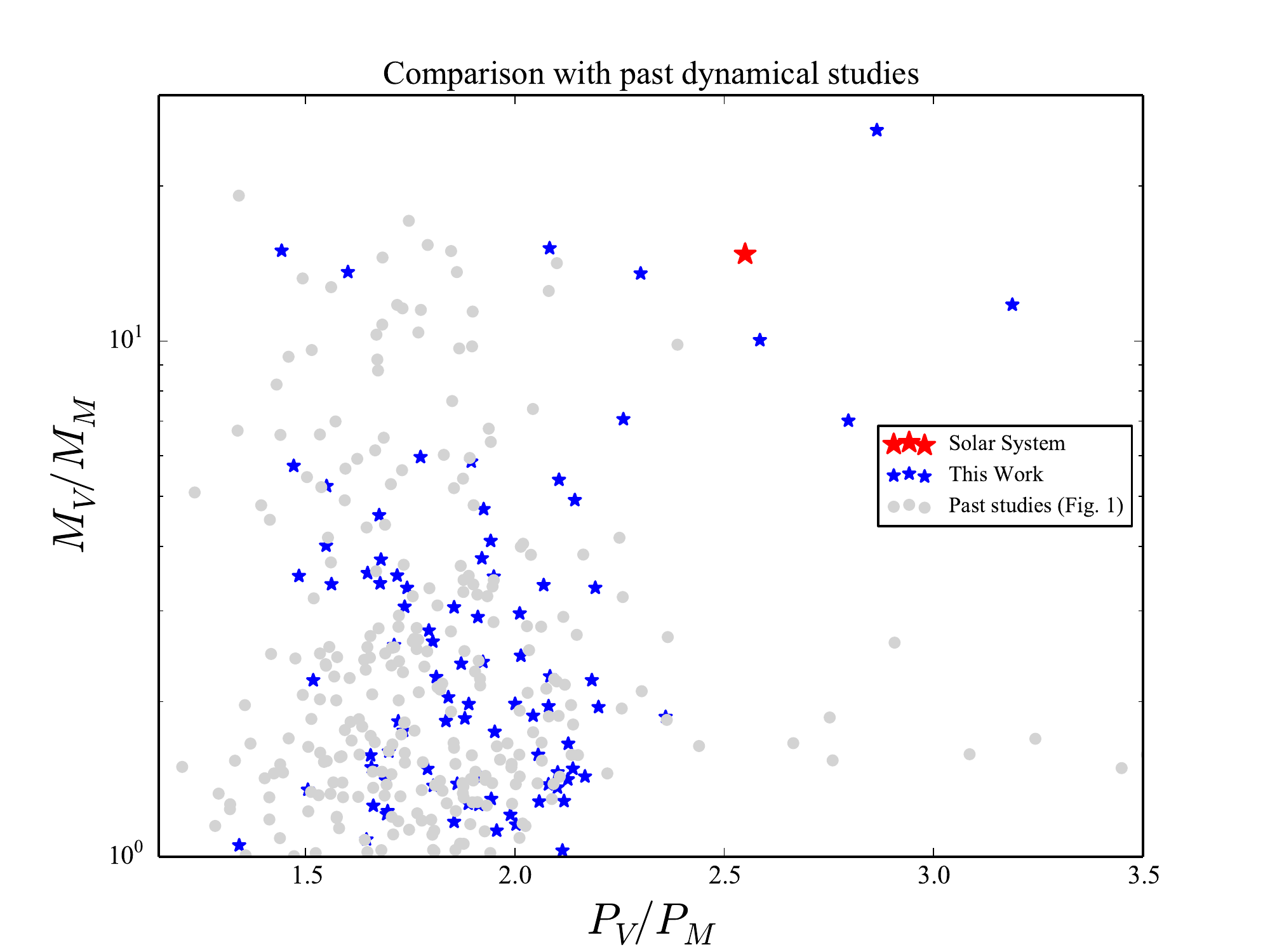}	
	\caption{The same as figure \ref{fig:new_fig} except, here, the systems formed in past dynamical studies \citep{chambers01,izidoro15,clement18,clement18_frag} are plotted in grey and criteria \textbf{A} satisfying realizations from this work are depicted with blue stars.  The red star denotes the solar system values of $P_{V}/P_{M}$ and $M_{V}/M_{M}$.}
	\label{fig:new_fig_tw}
\end{figure}

In figure \ref{fig:new_fig_tw} we revisit the parameter space of Venus-Mercury period and mass ratios depicted in figure \ref{fig:new_fig} by comparing the criterion \textbf{A} satisfying Mercury analogs generated in this paper, with those from past dynamical modeling efforts \citep{chambers01,izidoro15,clement18,clement18_frag}.  It is clear from the distribution of simulation outcomes that, while our proposed scenario substantially improves the efficiency of forming Mercury-Venus systems in broad strokes, the precise configuration remains a low-probability outcome in each dynamical model.  Encouragingly, the five or six most successful systems in figure \ref{fig:new_fig_tw} are derived from our present study.   Thus, our current manuscript serves as a proof-of-concept, and motivation for future study of mass-depleted inner disk components as a potentially viable means of consistently reproducing Mercury-like planets without invoking a low-probability giant impact.  Follow-on investigation should further probe the successful parameter space of disk structures uncovered in this paper, and work to incorporate and test the scenario's feasibility within the various proposed terrestrial evolutionary scenarios \citep[e.g.:][]{walsh11,levison15,bromley17,ray17sci,clement18}.

\section{Conclusions}

In this manuscript we presented a dynamical analysis of a scenario where Mercury forms directly within a mass-depleted, inner component of the terrestrial disk.  Our study follows the mold of previous investigations into the formation of Mars and the asteroid belt that varied the mass and radial concentration of planet forming material to determine the parameters that best reproduced the modern system \citep{chambers02,iz14,izidoro15}.  In general, we find our scenario markedly improves upon previous efforts to form Mercury in terms of its ability to consistently generate analogs of the modern Mercury-Venus system.  Additionally, we varied several disk parameters to determine which inner disk structures most efficiently formed adequate Mercury analogs.  Through this process, we conclude that our results are most sensitive to the total mass and radial extent of the inner disk, while the ratio of total embryo to planetesimal mass and the surface density profile (or slope) only mildly affect the final statistics.  Specifically, we determine that the planets' precise orbits are best reproduced when the mass depletion extends to just outside of Venus' modern orbit.  Furthermore, we find that moderate to larger total inner disk masses (0.1-0.25 $M_{\oplus}$) increase the likelihood of Mercury forming in dynamical isolation from Venus.  When the inner disk mass is too small, Mercury often survives as a stranded embryo, and is typically too close to Venus.  Finally, we note that our scenario provides a potent collisional environment that is capable of moderately modifying Mercury's composition through erosive collisions.  In successful simulations, Mercury's mantle is slowly eroded throughout the growth process and the final planet possess a massive iron core, in reasonable agreement with that of the actual planet.  The ejected fragments of mantle material combine with other embryos in the simulation that disproportionately tend to be excited onto orbits that collide with the Sun.

In spite of all efforts made, Mercury's precise orbit and mass remain difficult to explain in our scenario.  Specifically, the Venus-Mercury mass and orbital period ratios lie at the extreme of the distribution of outcomes generated in our simulations.  Nevertheless, our results are promising in terms of their consistent ability to generate Mercury-Venus pairs, and several of our simulations generate remarkably accurate inner solar system analogs. Thus, future investigations must thoroughly investigate the parameter space of initial conditions determined to be plausible in this manuscript, and simultaneously incorporate more detailed giant planet evolution and migration models.

\section*{Acknowledgments}

We thank Conel Alexander and an anonymous reviewer for insightful comments and input that greatly improved the manuscript.  We are also grateful to Andr\'{e} Izidoro, Seth Jacobson and Patryk Lykawka for graciously and promptly sharing data their from \citet{izidoro15}, \citet{jacobson14} and \citet{lykawka19} for the production of figures \ref{fig:new_fig} and \ref{fig:new_fig_tw}.  This work used the Extreme Science and Engineering Discovery Environment (XSEDE), which is supported by National Science Foundation grant number ACI-1548562. Specifically, it used the Comet system at the San Diego Supercomputing Center (SDSC) and the Bridges system, which is supported by NSF award number ACI-1445606, at the Pittsburgh Supercomputing Center \citep[PSC:][]{xsede}. Additional computation for the work described in this paper was supported by Carnegie Science's Scientific Computing Committee for High-Performance Computing (hpc.carnegiescience.edu).

\bibliographystyle{aasjournal}
\newcommand{\sci}{$Science$ }
\bibliography{merc2}
\end{document}